\documentclass[12pt,english]{article}
\usepackage[T1]{fontenc}
\usepackage[latin9]{inputenc}
\usepackage[letterpaper]{geometry}
\usepackage{amstext}
\usepackage{amsmath}
\usepackage{amssymb}
\usepackage{graphicx}
\usepackage{multirow}
\usepackage[round]{natbib}
\usepackage{authblk}
\usepackage{xr}

\begin{document}

\title{Temperatures in transient climates: improved methods for simulations with evolving temporal covariances}

\author[1]{Andrew Poppick}
\author[2]{David J. McInerney}
\author[3]{Elisabeth J. Moyer}
\author[1]{Michael L. Stein}
\affil[1]{Department of Statistics, University of Chicago}
\affil[2]{School of Civil, Environmental and Mining Engineering, University of Adelaide}
\affil[3]{Department of the Geophysical Sciences, University of Chicago}

\date{\vspace{-5ex}}

\maketitle

\begin{abstract}
Future climate change impacts depend on temperatures not only through changes in their means but also through changes in their variability. General circulation models (GCMs) predict changes in both means and variability; however, GCM output should not be used directly as simulations for impacts assessments because GCMs do not fully reproduce present-day temperature distributions. This paper addresses an ensuing need for simulations of future temperatures that combine both the observational record and GCM projections of changes in means and temporal covariances. Our perspective is that such simulations should be based on transforming observations to account for GCM projected changes, in contrast to methods that transform GCM output to account for discrepancies with observations.  Our methodology is designed for simulating transient (non-stationary) climates, which are evolving in response to changes in CO$_2$ concentrations (as is the Earth at present). This work builds on previously described methods for simulating equilibrium (stationary) climates. Since the proposed simulation relies on GCM projected changes in covariance, we describe a statistical model for the evolution of temporal covariances in a GCM under future forcing scenarios, and apply this model to an ensemble of runs from one GCM, CCSM3. We find that, at least in CCSM3, changes in the local covariance structure can be explained as a function of the regional mean change in temperature and the rate of change of warming. This feature means that the statistical model can be used to emulate the evolving covariance structure of GCM temperatures under scenarios for which the GCM has not been run. When combined with an emulator for mean temperature, our methodology can simulate evolving temperatures under such scenarios, in a way that accounts for projections of changes while still retaining fidelity with the observational record. The emulator for variability changes is also of interest on its own as a summary of GCM projections of variability changes.
\end{abstract}

\section{Introduction}
Assessing the potential impacts of future climate change on areas of societal interest, such as agriculture and public heath, requires an understanding of how climate features important to those areas are expected to change. Impacts often depend on more than just the response of global or even local mean temperatures to greenhouse gas forcing. Many agricultural crops, for example, are highly sensitive to even brief periods of stress temperatures, particularly at certain times of the growing cycle, so crop yields can be strongly affected by changes in temperature variability even in the absence of a change in mean (e.g., \cite{wheeler}). In part because of examples like this, the climate and impacts communities have been interested in understanding changes in temperature variability in future climates.

Potential future changes in temperature variability are not yet well understood. By its third assessment report, the Intergovernmental Panel on Climate Change (IPCC) stated that there was some empirical evidence for a decrease in variability at intra-annual timescales, but sparse evidence for changes in inter-annual variability~\citep{tar}. More recent studies have not produced more definitive conclusions, with results apparently depending on specific definitions of variability and timescale as well as on the region being studied~\citep{AR4Variability}. (The most recent report,~\cite{AR5}, frames variability changes in the context of extreme events, which are not a subject of this work.) One example of a physical mechanism that might explain variability changes at intra-annual timescales in a particular region is that changes in the polar jet stream can produce more persistent weather patterns over, for example, North America  (e.g., \cite{francis}), but the mechanism and even detection of these changes remains controversial~\citep{screen, barnes}. Implicit in the broader discussion about variability is that variability changes can differ by timescale of variation or geographic location. Because specific impacts will depend on timescale and geographic location, methods for assessing changes should be able to make such distinctions. That is, understanding projected changes in variability relevant to impacts is a problem of understanding the changes in covariance structure of a spatial-temporal field that is evolving in time.

Beyond empirical studies, the primary tools used to understand and project changes in the distribution of climate variables are atmosphere-ocean general circulation models (GCMs). GCMs are deterministic, physical models that are used to generate runs of modeled climate under, for example, varying forcing scenarios. While GCMs are deterministic, the climate system being modeled is chaotic and so GCM realizations under the same forcing scenario but with different initial conditions will behave as if they were statistically independent. Summarizing the statistical properties of GCM predictions under different forcing scenarios is a challenge on its own.

That said, GCM runs alone are often not sufficient as inputs for impacts assessments, which may require realistic simulations from the full distribution of the Earth's temperatures. It is well understood that GCMs somewhat misrepresent observed temperature distributions under present conditions: regional mean temperatures may differ from observations by several degrees, and studies have noted discrepancies between higher order moments of the modeled and observed climate distributions (\cite{AR5} and references therein). On the other hand, GCMs are assumed to produce informative projections of, for example, future changes in mean temperatures due to changes in greenhouse gas forcing: the underlying physics are relatively realistic, and GCMs are able to reproduce observed temperature trends in historical forcing runs (e.g., \cite{AR5}). Projections of variability changes also have many consistent features across different GCMs, although current studies do not address changes in full covariance structures (e.g., \cite{schneider2015,holmes2015}). Impacts assessments researchers have therefore recognized a need to understand not only how temperature distributions are changing in GCM runs, but also how to combine those projections with the observational record to produce simulations of temperatures that more likely follow the distribution of real future temperatures. 

There are two popular classes of approaches for generating such simulations: those that modify GCM output to account for model-observation discrepancies (model-driven procedures), and those that modify observational data to account for changes projected by GCMs (observation-driven procedures); see, for example, \cite{ho} and \cite{hawkins} for reviews of common strategies. The most basic model-driven procedure is simple ``bias correction'', where the difference in mean between observed temperatures and those in historical GCM realizations is assumed constant over time, and the estimated bias is subtracted from future GCM runs. The most basic observation-driven procedure is the Delta method\footnote{The meaning of the term ``Delta method'' in the geosciences, and in this work, is distinct from its typical use in the statistics literature for methods that employ Taylor expansions to derive asymptotic approximations of properties of functions of random variables.}, where, by contrast, changes in mean temperature are estimated by comparing GCM future realizations with those under historical forcing, and then this trend is added to the observational data. Both approaches implicitly assume that GCMs correctly capture changes in mean temperature; however, they can result in temperature simulations that have very different higher-order characteristics.

\begin{figure}[!t]
\begin{center}
\includegraphics[scale = 0.3, trim = 5cm 2cm 0cm 0cm]{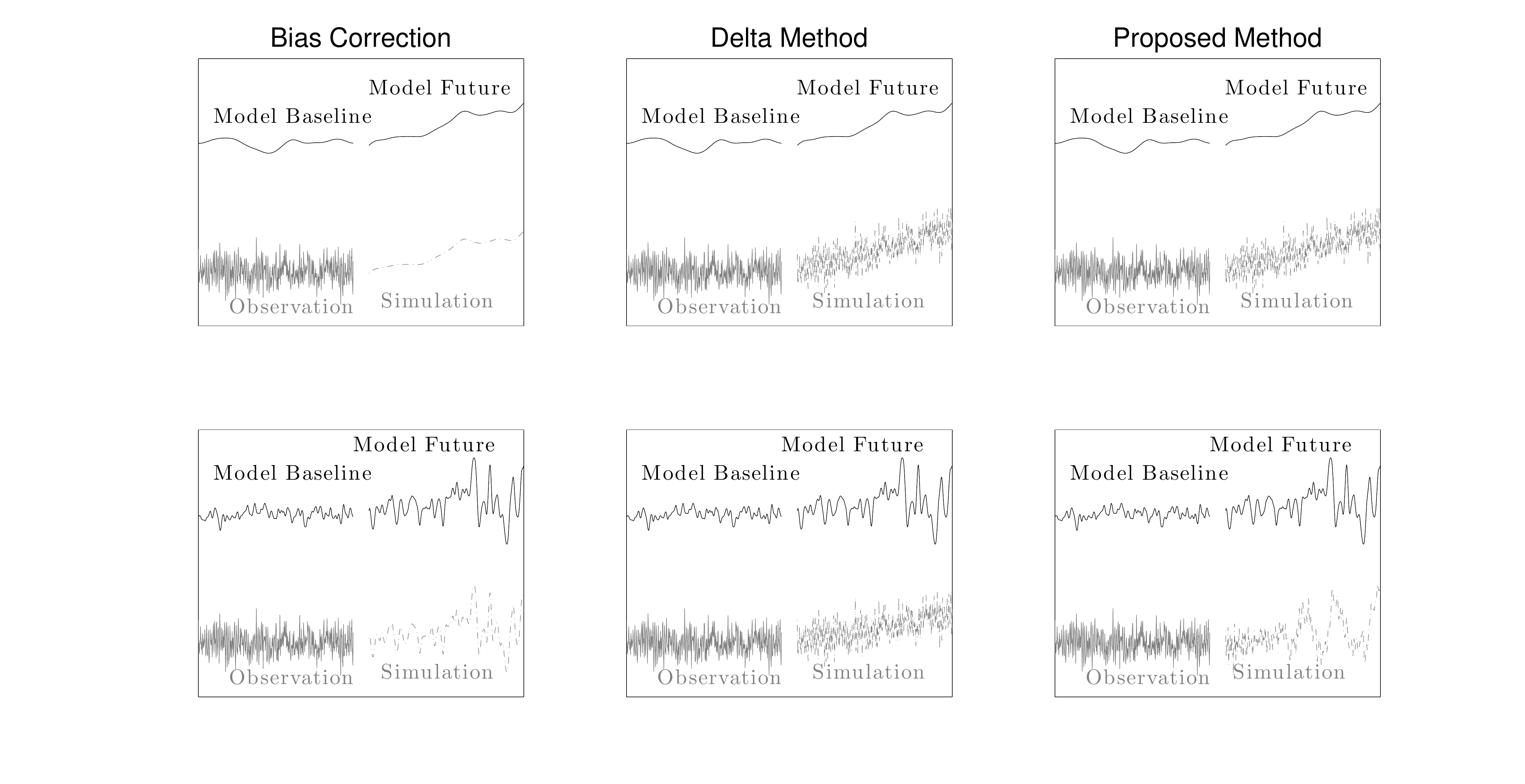}
\caption{Cartoon illustration comparing strategies for simulating temperatures that combine information from a model and the observational record. Columns compare simple bias correction (left), the Delta method (center), and our proposed method (right). Top row, the model predicts changes in mean temperature but no changes in variability; in this case, our proposed method is equivalent to the Delta method. Bottom row, the model predicts changes in both mean and covariance. Simple bias correction does not retain the higher order properties of the observations, whereas the Delta method does not account for model changes in covariance; our proposed method does both.}
\label{fig:illust}
\end{center}
\end{figure}

An appealing property of observation-driven procedures like the Delta method is that they preserve attributes of the observations that are not explicitly accounted for in the simulation procedure, a property not shared by model-driven procedures. Figure~\ref{fig:illust}, top row, provides a cartoon illustration of this difference between the two approaches. Here, the cartoon model predicts changes in mean but badly misrepresents the mean and covariance structure of the  observations. In such a setting, simple bias correction will maintain the model's misrepresented covariance structure, while the Delta method yields a more realistic simulation (see~\cite{hawkins} for a less idealized example). More complicated versions of bias correction exist that attempt to correct for higher-order discrepancies between models and observations. Some correct for discrepancies in marginal distributions (e.g., \cite{wood}), while others additionally attempt to correct rank correlation structures and inter-variable dependence structures (e.g., \cite{piani,vrac}). While such methods are more sophisticated than the simple bias correction illustrated in Figure~\ref{fig:illust}, they too will leave intact discrepancies between the model and observations not accounted for in the correction procedure. If impacts assessments require realistic simulations from the joint distribution of temperatures across space and time, our perspective is that this objective is more easily met by observation-driven methods.

Other routinely used simulation methods exist that do not fall as neatly within the model-driven/observation-driven dichotomy. For example, in simulations produced by stochastic weather generators~\citep{semenov1997,wilks1999}, the observations are replaced with synthetic data drawn from a stochastic model meant to mimic the distribution of the observations. The stochastic weather generator can then be modified to account for changes predicted by a climate model. The drawback of this approach is that a statistical model for the observations is required in addition to a model for GCM projected changes, whereas observation-driven methods only require the latter. Synthetically generated observations will be less realistic than the observations themselves. Other related methods in the statistics literature also attempt to use statistical models to blend information from observations with climate models to produce future simulations (e.g.,~\cite{salazar2011}), but the proposed statistical models make very strong assumptions on the spatiotemporal distribution of the observations and the climate model realizations. Especially when projected changes from the historical climate are not very large, simulation methods should preserve features of the observed climate where possible. We therefore view observation-driven methods like the Delta method as likely to produce more realistic simulations than these methods as well.

Current observational data products are not themselves without some uncertainties that will affect all of the simulation methods we have described, since they all rely on observational data in some way. For example, uncertainties emerge from the inhomogeneous distribution of observation locations, measurement error of a potentially time-varying nature, and interpolation schemes used both to produce gridded data products and to downscale or upscale from one spatial grid to another. The last point highlights the important issue of spatial change of support and misalignment~\citep{gotway}: some raw observational data are point-referenced (e.g., station data), whereas others represent area averages (e.g., satellite data), so care must be taken to appropriately combine these data to produce gridded products. Some observational data products are  distributed as ensembles in an attempt to explicitly account for uncertainties (e.g., \cite{morice2012}). \cite{AR5}, Chapter 2 (Box 2.1 and elsewhere), and references therein contain discussions of uncertainty in the observational record. In this work, we will illustrate our method using one observational product, but as with all existing simulation methods that use observations, the method we will propose can be used with any and all available data products.

While we have advocated for observation-driven simulation methods, an important limitation of the observation-driven Delta method described above, is that it does not account for changes in variability (Figure~\ref{fig:illust}, bottom row). Some extensions of the Delta method account for changes in marginal variance projected by a GCM (e.g., see again~\cite{hawkins}), but, again, since variability changes need not be uniform across all timescales of variation, changes in marginal variance are not a complete summary of GCM projected changes in variability. \cite{leeds} introduced an extension of the Delta method that does account for changes in the full temporal covariance structure projected by a GCM, but their approach is applicable only for equilibrated climates, in which temperatures (after preprocessing for seasonality) can be assumed to be stationary in time. The Earth's climate, however, is and will continue to be in a transient state, in which temperatures will by definition be nonstationary in time. There is therefore an outstanding need for methods both to characterize changes in covariance in transient, nonstationary climates, and to simulate temperatures in such climates.

In this work, we build on the work described in~\cite{leeds} to develop a methodology for generating observation-driven simulations of temperatures in future, transient climates that account for transient changes in both means and temporal covariances. In Figure~\ref{fig:illust}, bottom row, our proposed method, unlike simple bias correction or the Delta method, both accounts for the relevant changes projected by the cartoon model and retains other distributional properties of the observations. Our method reduces to the Delta method in the case that the model predicts no changes in variability, and reduces to the method in~\cite{leeds} if the past and future climates are both in equilibrium. Since such a simulation uses projected changes in covariances from a GCM, our methodology must provide a way of modeling and estimating these changes in transient GCM runs. The transient, nonstationary setting adds substantial challenges, and so the statistical modeling of changes in covariance in transient GCM runs is a primary focus of this paper.

As a final complication, since GCMs are extremely computationally intensive, it is not possible to run a GCM under every scenario relevant for impacts assessments. In the absence of a run for a scenario of interest, impacts modelers may instead rely on a GCM emulator, a simpler procedure that produces, for example, mean temperatures that mimic what the GCM would have produced had it been run. Our framework for simulations can use emulated rather than true GCM projections. For methods that emulate mean temperatures over forcing scenarios, see \cite{castruccio} and references therein, and the literature stemming from the pattern scaling method of ~\cite{santer}. Much of the literature on climate model emulation has focused not on emulating model output across different forcing scenarios, but rather on emulating output with differing values of key climate model parameters (often for the purpose of selecting values of those parameters) (e.g., \cite{chang2014,rougier2009,sanso2008,sanso2009,sham2012,williamson2013} and others). While the statistical concerns related to emulating climate models in parameter space are somewhat different from those of emulating in scenario space, the general idea remains the same: that one may use available climate model runs to infer properties of a run that has not been produced. In our case, we require an emulator for the GCM changes in covariance in addition to a mean emulator. Our proposed statistical model can be used for this purpose, allowing our observation-driven procedure to simulate future temperatures in a potentially wide range of forcing scenarios.

The remainder of this article is organized as follows. In Section 2, we motivate and describe our procedure for observation-driven simulations of future temperatures in transient climates that accounts for projected changes in both means and temporal covariances. In Section 3, we describe a GCM ensemble that we use to illustrate our methodology. In Section 4, we describe a statistical model for the changes in temporal covariances observed in this GCM ensemble that can be used as an emulator for these changes; we also discuss the estimation of this statistical model. In Section 5, we discuss results, both in terms of the GCM projections and the corresponding simulations, as well as the quality of our model in emulating the GCM projections. In Section 6, we give some concluding remarks and highlight areas for future research.

\section{Observation-driven simulations of temperatures in future transient climates}
Our goal is to provide a simulation of future temperatures in a transient climate under a known forcing scenario. In light of the preceding discussion, this simulation should reflect knowledge of the changes in the mean and covariance structure of future temperatures under that scenario, but should otherwise preserve properties of the observed temperature record. 

Our proposed procedure is motivated by an idealization of the problem, supposing that the future changes in mean and temporal covariance structure are known. Following this motivation, we describe some modifications to the proposed procedure that we argue make the procedure more useful in practical settings when changes in mean and covariance must be estimated from, for example, GCM runs.

\subsection{Idealization}
Consider a family of multivariate (i.e., spatially referenced) Gaussian time series, $z^{(s)}_{l}(t)$ at times $t = ...,-1, 0, 1, ...$ and locations $l = 1,...,L$, indexed by $s \in \mathcal{S}$, some set of scenarios. Write $\mu^{(s)}_{l}(t)$ for the unknown mean of $z^{(s)}_{l}(t)$ and assume that at each location, $z^{(s)}_{l}(t)$ has an unknown evolutionary spectrum, $a^{(s)}_l(t,\omega)$; for details on processes with evolutionary spectra, see~\cite{priestley}. Processes with temporally varying covariance structures in general have been discussed extensively in the literature. An overview of a theoretical framework for understanding locally stationary processes, closely tied to the Priestley model, can be found in~\cite{dahlhaus2012} and references therein. Our focus in this paper is on spectral methods because we view evolutionary spectra to be an intuitive way to characterize time-varying covariances and because the process's corresponding spectral representation has useful implications for our simulation procedure. Most importantly, $z^{(s)}_{l}(t)$ has, at each location and for each $s$, the spectral representation
$$z^{(s)}_{l}(t) =\mu_{l}^{(s)}(t) + \int_{-\pi}^{\pi} e^{i \omega t} \sqrt{a_l^{(s)}(t,\omega)} \, d\xi_l^{(s)}(\omega),$$
where $\xi_l^{(s)}(\omega)$ is a mean zero process with orthogonal increments and unit variance; that is, $E[d\xi_l^{(s)}(\omega)d\xi_l^{(s)}(\omega)^*] =d\omega$ and $E[d\xi_l^{(s)}(\omega)d\xi_l^{(s)}(\omega')^*]=0$ for $\omega\ne\omega'$, with $^*$ denoting the complex conjugate. Here and throughout this paper, we restrict our attention to nonstationary processes with evolutionary spectral representations whose transfer functions, $\sqrt{a_l^{(s)}}$, are real and positive.

Suppose that we observe the time series under one scenario, $z^{(0)}_{l}(t)$, for times $t = 1,...,N_0$. Call $z^{(0)}_{l}(t)$ the observed time series; given the observed time series, we would like to generate a simulation of the same length as the observations, but approximately equal in distribution to an unobserved time series, $z^{(s)}_{l}(t)$, for a given $s$. Since both $z^{(0)}_{l}(t)$ and $z^{(s)}_{l}(t)$ are Gaussian, there is a class of affine transformations of $z^{(0)}_{l}(t)$ that is equal in distribution to $z^{(s)}_{l}(t)$; indeed, writing $\Sigma_l^{(0)}$ and $\Sigma_l^{(s)}$ for the covariance matrices of the observed and unobserved time series at location $l$, we have that (marginally, at each location) 
$$\mathbf{z}^{(s)}_{l} =_\mathcal{D} \boldsymbol{\mu}^{(s)}_{l} + (\Sigma_{l}^{(s)})^{1/2} (\Sigma_{l}^{(0)})^{-1/2}(\mathbf{z}^{(0)}_{l} - \boldsymbol{\mu}^{(0)}_{l})$$
for any matrix square root, where $\mathbf{x}$ denotes the vector with entries $x(t)$. While it is not immediately obvious that this fact is helpful, since the means and covariances of the two time series are unknown and at least for the unobserved time series cannot be directly estimated, we will describe a setting in which it is possible to compute this transformation (approximately) without fully knowing the means and covariances of the two time series.

The covariances of the observed and unobserved time series may be written as
\begin{eqnarray*}
(\Sigma^{(0)}_{l})_{t,t'} & = & \int_{-\pi}^{\pi} e^{i \omega (t-t')} \sqrt{a_l^{(0)}(t,\omega)a_l^{(0)}(t',\omega)^*} \, d\omega,\text{ and} \\
(\Sigma^{(s)}_{l})_{t,t'} & =  & \int_{-\pi}^{\pi} e^{i \omega (t-t')} \sqrt{a_l^{(s)}(t,\omega)a_l^{(s)}(t',\omega)^*} \, d\omega,
\end{eqnarray*}
which follows immediately from the processes' spectral representations.~\cite{guinness} showed that, under some regularity conditions on the evolutionary spectra, these matrices can be approximated as
\begin{eqnarray*}
\Sigma_l^{(0)} & \approx & C_{N_0}\left(\sqrt{a_l^{(0)}}\right) C_{N_0}\left(\sqrt{a_l^{(0)}}\right)^H, \text{ and } \\
\Sigma_l^{(s)} & \approx & C_{N_0}\left(\sqrt{a_l^{(s)}}\right) C_{N_0}\left(\sqrt{a_l^{(s)}}\right)^H
\end{eqnarray*}
where $H$ denotes the conjugate transpose and generically for some function $A(t,\omega)$ in time and frequency, $C_N(A)$ is the $N\times N$ matrix with entries
$$C_N(A)_{t,j} = \sqrt{\frac{2\pi}{N}}\, A(t,2\pi (j-1)/N)e^{2\pi i (j-1) t/N}$$
for $t,j = 1,...,N$. (In the setting where $A$ is constant in time, $C_N(A)$ is the inverse discrete time Fourier transform scaled by $A$ and the result is the well-known result that the discrete time Fourier transform approximately diagonalizes the covariance matrix for a stationary time series observed at evenly-spaced intervals.) The following transformation of $z_{l}^{(0)}(t)$ is therefore, marginally at each location, $l$, approximately equal in distribution to $z_{l}^{(s)}(t)$:
\begin{equation}
\label{eq:simulation1}
\mathbf{z}_{l}^{(s,0)} = \boldsymbol{\mu}_{l}^{(s)} +  C_{N_0}\left(\sqrt{a_l^{(s)}}\right)  C_{N_0}\left(\sqrt{a_l^{(0)}}\right)^{-1} (\mathbf{z}_{l}^{(0)} - \boldsymbol{\mu}_{l}^{(0)}).
\end{equation}

The crucial observation, however, is that~\eqref{eq:simulation1} can be computed exactly without fully knowing the means and covariances of the observed and unobserved time series.  Indeed, suppose that what we are given are not the means and evolutionary spectra of the processes themselves, but some other substitute set of functions $\tilde{\mu}^{(s)}_{l}(t)$ and $\tilde{a}_l^{(s)}(t,\omega)$ satisfying, for each scenario $s$ and at each location $l$,
\begin{equation}
\label{eq:assumptions}
\mu^{(s)}_{l}(t)  - \tilde{\mu}^{(s)}_{l}(t) = c_l, \text{ and } \frac{a_l^{(s)}(t,\omega)}{\tilde{a}_l^{(s)}(t,\omega)} = k_l(\omega),
\end{equation}
for some unknown constant $c_l$ and some unknown function $k_l(\omega)$ that is constant in time. This situation is analogous to our actual predicament, where GCM runs are assumed to be more informative about changes than absolute levels; indeed, one consequence of the assumptions~\eqref{eq:assumptions} is that the substitute means and evolutionary spectra change in the same way as their true counterparts, so, for instance, we may write
$$\Delta_{l}^{(s,0)}(t) \equiv \mu^{(s)}_{l}(t) - \mu^{(0)}_{l}(t)  = \tilde{\mu}^{(s)}_{l}(t) - \tilde{\mu}^{(0)}_{l}(t), \text{ and }  $$
$$\rho_l^{(s,0)}(t,\omega) \equiv \frac{a_l^{(s)}(t,\omega)}{a_l^{(0)}(t,\omega)} = \frac{\tilde{a}_l^{(s)}(t,\omega)}{\tilde{a}_l^{(0)}(t,\omega)},$$
for the known changes in means and covariance structures. The assumption that GCM mean temperatures are off by a constant compared to real temperatures is essentially the assumption underlying both simple bias correction and the Delta method as described in Section 1, and we view the assumption on the evolutionary spectra as a natural extension to covariances; all existing simulation methods that we are aware of implicitly or explicitly make the same or similar assumptions (except the simple Delta method, which assumes no changes in variability at all). 

Under these assumptions,~\eqref{eq:simulation1} may be rewritten as
\begin{equation}
\label{eq:simulation2}
\mathbf{z}_{l}^{(s,0)} = \boldsymbol{\mu}_{l}^{(0)} + \boldsymbol{\Delta}_{l}^{(s,0)}+  C_{N_0}\left(\sqrt{\tilde{a}_l^{(s)}}\right)  C_{N_0}\left(\sqrt{\tilde{a}_l^{(0)}}\right)^{-1} (\mathbf{z}_{l}^{(0)} - \boldsymbol{\mu}_{l}^{(0)}),
\end{equation}
where, to reiterate,~\eqref{eq:simulation1} and~\eqref{eq:simulation2} are equal under the assumptions~\eqref{eq:assumptions} because $C_{N_0}\left(\sqrt{\tilde{a}_l^{(s)}}\right) = C_{N_0}\left(\sqrt{a_l^{(s)}}\right) \mbox{diag}(1/k_l)$, where $\mbox{diag}(1/k_l)$ is the diagonal matrix with entries $1/k_l(\omega_j)$, so 
\begin{equation*}
C_{N_0}\left(\sqrt{\tilde{a}_l^{(s)}}\right)  C_{N_0}\left(\sqrt{\tilde{a}_l^{(0)}}\right)^{-1}  = C_{N_0}\left(\sqrt{a_l^{(s)}}\right) C_{N_0}\left(\sqrt{a_l^{(0)}}\right)^{-1}.
 \end{equation*}
In light of~\eqref{eq:simulation2}, our proposed simulation can be computed as long as one knows (or, more realistically, can estimate) just the mean of the observed time series as well as the substitute evolutionary spectra and changes in mean. The procedure described by~\eqref{eq:simulation2} is what is illustrated in the bottom right panel of Figure~\ref{fig:illust}.

In the case that there are no changes in covariance structure, so $a_l^{(s)}(t,\omega) = a_l^{(0)}(t,\omega)$ and the same for their substitute counterparts, the simulation procedure~\eqref{eq:simulation2} is equivalent to the Delta method as described in Section 1. In the case that both $a_l^{(s)}(t,\omega)$ and $a_l^{(0)}(t,\omega)$ are constant in time, so the de-meaned time series are stationary, the procedure is the same as that described in~\cite{leeds}. Our proposal is therefore a generalization of those two procedures, describing an observation-driven simulation that transforms one observed time series (possibly itself from a transient climate) to a simulation under a new (future, transient) scenario.

\subsection{Practical modifications to idealized procedure}
In practice, we do not actually know even substitute versions of the future changes in mean and covariance structure, so the procedure we have described in the preceding section must be modified to be made useful.

A key assumption underlying our methodology is that GCM runs are informative about at least some aspect of the changes in mean and covariance structure of the real temperatures; however, it need not be true that the assumptions~\eqref{eq:assumptions} will be satisfied by taking $\mu^{(s)}_{l}(t)$ and $a_l^{(s)}(t,\omega)$ to be the means and evolutionary spectra corresponding to real temperatures and taking $\tilde{\mu}^{(s)}_{l}(t)$ and $\tilde{a}_l^{(s)}(t,\omega)$ to be those corresponding to temperatures under GCM runs.  One possible objection to these assumptions is that both the observations and the GCM runs will exhibit nonstationarity in mean and variance due to seasonality, and it is at least plausible that the GCM representation of these seasonal cycles will differ from that of the observations. \cite{leeds} argued that the seasonality can be reasonably represented as a uniformly modulated process (see~\cite{priestley}) plus a mean seasonal cycle. That is, writing $T^{(s)}_{l}(t)$ for the true temperatures at time $t$ and location $l$ in scenario $s$, and $d$ for the day of the year, we assume that
$$T^{(s)}_{l}(t) = \mu^{(s)}_{l}(t) + m^{(T)}_{l}(d) + D^{(T)}_{l}(d) (z^{(s)}_{l}(t) - \mu^{(s)}_{l}(t))$$
where $m^{(T)}_{l}(d)$ and $D^{(T)}_{l}(d)$ represent seasonal cycles in means and marginal variances, and $z^{(s)}_{l}(t)$ has mean $\mu^{(s)}_{l}(t)$ and evolutionary spectrum $a^{(s)}(t,\omega)$ as above; assume a similar form for the GCM runs. (In the following, we will allow $\mu^{(s)}_{l}(t)$ to reflect changes in the mean seasonal cycle from $m^{(T)}_{l}(d)$ but for simplicity will assume that $a^{(s)}(t,\omega)$ has no seasonal structure; see Section 3.1 for details.) We estimate the seasonal cycles in mean and variability in the observations according to the methods described in~\cite{leeds}; the mean seasonal cycle is modeled with the first ten annual harmonics and estimated via least squares, whereas the annual cycle in variability is estimated by a normalized moving average of windowed variances that has been averaged across years. We will assume that~\eqref{eq:assumptions} is reasonable taking $\mu^{(s)}_{l}(t)$, $a^{(s)}(t,\omega)$, $\tilde{\mu}^{(s)}_{l}(t)$, and $\tilde{a}^{(s)}(t,\omega)$ to be the means and evolutionary spectra of the de-seasonalized components of real and GCM temperatures.

Even assuming that~\eqref{eq:assumptions} holds for the de-seasonalized components of observed and GCM temperatures, what we are given are the real and GCM temperatures themselves, not their means and evolutionary spectra. As such, the quantities necessary to compute~\eqref{eq:simulation2} must be estimated using the available data. While it would be possible to estimate the corresponding evolutionary spectra from the GCM runs directly, note that~\eqref{eq:simulation2} works for any substitute function in time and frequency that, for each frequency, is proportional to the true evolutionary spectra at all times. If such a function must be estimated, there is presumably a statistical advantage to estimating a function that is relatively flat across frequencies. In GCM experiments, it is fairly typical to have a control run under an equilibrated (often preindustrial) climate, in which at least the de-seasonalized temperatures can be viewed as a stationary process. Writing $s=B$ for this equilibrated scenario, and $\tilde{a}_l^{(B)}(\omega) \equiv \tilde{a}_l^{(B)}(t,\omega)$ for the corresponding spectral density of the de-seasonalized component of the equilibrated GCM temperatures, then if~\eqref{eq:assumptions} holds, it will also be true that we can write
$$\rho^{(s,B)}_l(t,\omega) \equiv \frac{a^{(s)}(t,\omega)}{a^{(B)}(\omega)} = \frac{\tilde{a}^{(s)}(t,\omega)}{\tilde{a}^{(B)}(\omega)},$$
in which case $a^{(s)}(t,\omega) / \rho^{(s,B)}_l(t,\omega) = a^{(B)}(\omega)$ and~\eqref{eq:assumptions} still holds if one replaces $\tilde{a}^{(s)}(t,\omega)$ with $\rho^{(s,B)}_l(t,\omega)$. Moreover, we expect that the functions $\rho^{(s,B)}_l(t,\omega)$ will be much flatter than the functions $\tilde{a}^{(s)}_l(t,\omega)$ over the range of scenarios considered reasonable, so we expect that there should be some advantage in estimating these ratios rather than the evolutionary spectra themselves from the GCM runs.

Writing $T^{(s,0)}_{l}(t)$ for our simulation of the true temperatures under scenario $s$, our proposed procedure is therefore
\begin{eqnarray}
\label{eq:simulation}
\mathbf{T}^{(s,0)}_{l} &  =  & \hat{\boldsymbol{\mu}}^{(0)}_{l} +  \hat{\mathbf{m}}^{(T)}_{l} + \hat{\boldsymbol{\Delta}}_{l}^{(s,0)}  \\ 
& & + \mbox{diag}(\hat{\mathbf{D}}^{(T)}_{l}) C_{N_0}\left(\sqrt{\hat{\rho}^{(s,B)}}\right)C^{-1}_{N_0}\left(\sqrt{\hat{\rho}^{(0,B)}}\right) (\hat{\mathbf{z}}_{l}^{(0)} - \hat{\boldsymbol{\mu}}^{(0)}_{l}) \notag,
\end{eqnarray}
where $\hat{x}$ generically represents an estimate of the quantity $x$. In words, the procedure is, in order, (i) estimate and remove seasonality and mean trend in the observational record; (ii) estimate the future changes in mean and marginal spectra using an ensemble of GCM runs; (iii) de-correlate the de-trended and de-seasonalized observations using the estimated substitute function, $\hat{\rho}^{(0,B)}$, obtained in step (ii);  (iv) apply the estimated changes in spectra, $\hat{\rho}^{(s,B)}$, to the decorrelated data and invert the transformation in (iii);  and (v) replace the seasonal cycles in means and variability, and add the new changes in mean.

The practicality of~\eqref{eq:simulation} depends both on our ability to obtain good estimates of all of the involved quantities and on our ability to compute the simulation efficiently.~\cite{castruccio} and~\cite{leeds} collectively describe methods that essentially can be used to estimate all of the necessary quantities in the procedure except, crucially, the changes in evolutionary spectra. In the following, we will discuss modeling and estimating these functions from an ensemble of GCM runs. The particular statistical model we develop allows for efficient computation of the simulation.

\section{Description of GCM ensemble}
\begin{figure}[!t]
\begin{center}
\includegraphics[scale = 0.25]{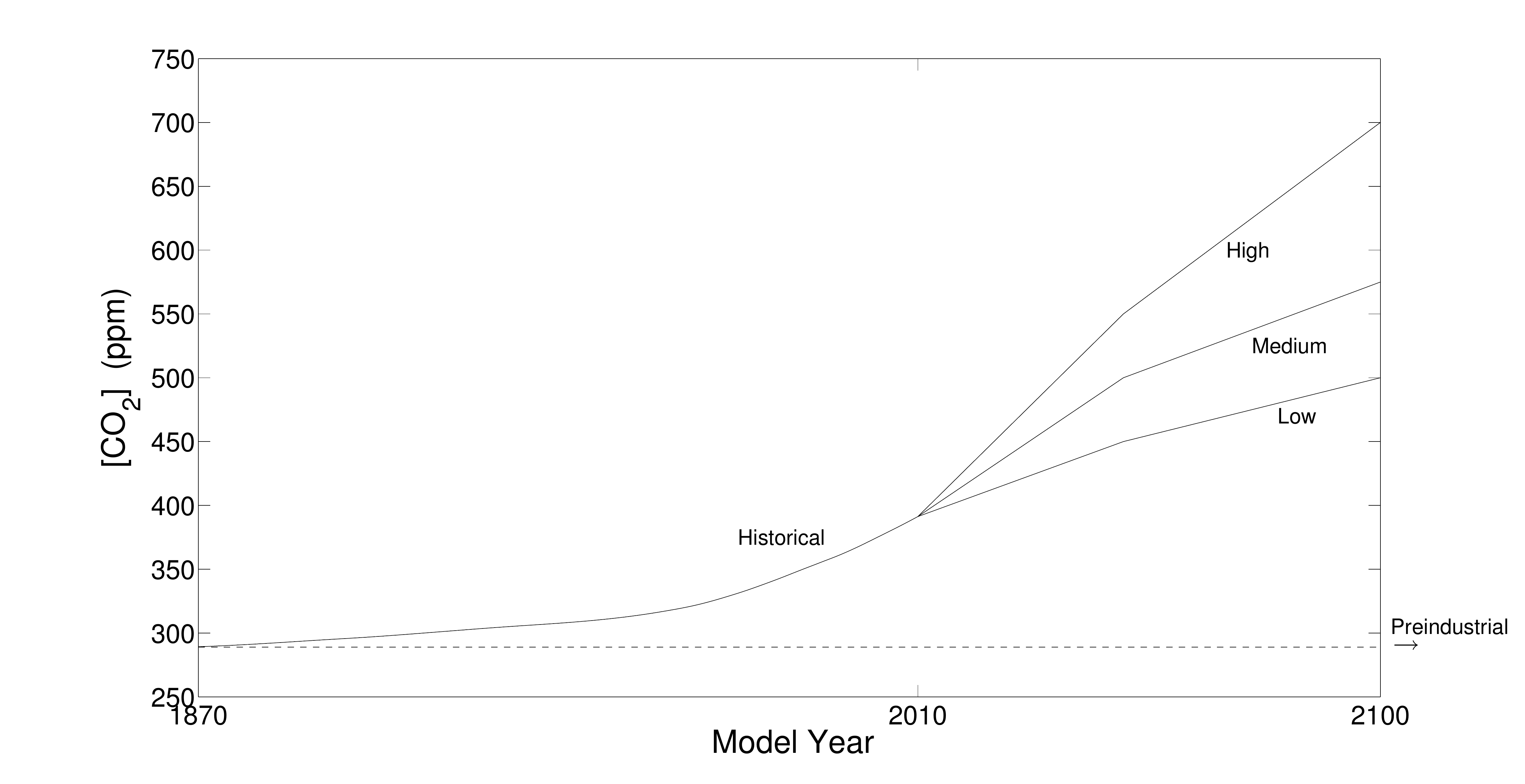}
\caption{GCM ensemble $[\mbox{CO}_2]$ trajectories. The historical scenario begins in 1870 and follows observed $[\mbox{CO}_2]$ until 2010, after which it branches into the three future scenarios increasing at different rates until 2100. The preindustrial run maintains 1870-level $[\mbox{CO}_2]$ until the year 4600, but we use only the last 1,500 years of that run. The ensemble includes eight realizations each under the historical, high, and low scenarios, five under the medium scenario, and one under preindustrial $[\mbox{CO}_2]$.}
\label{fig:CO2}
\end{center}
\end{figure}

We study changes in the distribution of daily temperatures in an ensemble of GCM runs made with the Community Climate System Model Version 3 (CCSM3) \citep{CCSM3,yeager} at T31 atmospheric resolution (a $48 \times 96$ grid with a resolution of approximately $3.75^\circ \times 3.75^\circ$), and a $3^\circ$ resolution for oceans. All runs require a long spin up; the realizations  in our ensemble are initialized successively at ten-year intervals of the NCAR b30.048 preindustrial control run. Each transient realization is then forced by historical CO$_2$ concentrations (herein, [CO$_2$]) from years 1870-2010, at which point the ensemble branches into three future increasing [CO$_2$] scenarios for the years 2010-2100, which we name the ``high'', ``medium'', and ``low'' concentration scenarios (Figure~\ref{fig:CO2}).  For each scenario, we have a modest number of realizations (eight realizations from the historical, high, and low scenarios, and five realizations from the medium scenario), so the transient ensemble consists of about 1.1 million observations at each grid cell, or about 5 billion observations in total. As is typical, we will assume throughout that the ensemble members can be viewed as statistically independent realizations due to the system's sensitivity to initial conditions.

The focus of our investigation is on changes in temporal covariance structure in transient (nonstationary) runs of the GCM, but for the reasons described in Section 2.2, it is helpful to have a representation of the model's climate in a baseline, equilibrated state. For this purpose, we use a single, long run forced under preindustrial [CO$_2$] (289 ppm) for an additional 2,800 years past the control run initialization to ensure that the run is fully equilibrated, from which we use the last 1,500 years, or about 0.5 million days, for a total of about 2.5 billion observations under pre-industrial [CO$_2$]. 

In the following, we will index the members of the GCM ensemble by their [CO$_2$] scenario, $s \in \{B, 0, H, M, L\}$, denoting, respectively, the baseline preindustrial, historical, high, medium, and low scenarios.

\subsection{Data preprocessing}The primary inferential aim of this work is to obtain estimates of $\rho^{(s,B)}_l(t,\omega)$, the changes in marginal evolutionary spectra of the de-seasonalized component of daily temperatures in the GCM under scenario $s$ compared to the preindustrial climate. The GCM runs have, accordingly, been preprocessed to remove means and seasonal cycles of variability.

As with the observed temperatures in Section 2, we represent temperatures in the preindustrial run at each grid cell as a uniformly modulated process plus a mean seasonal cycle and retain the stationary component of this process. That is, write $y_{l}^{(B)}(t)$ for the temperature in the raw, equilibrated preindustrial GCM run at time $t$ and location $l$, and again write $d$ for the day of the year (the GCM does not have leap years, so $d\in\{1,...,365\}$). We represent these as
$$ y_{l}^{(B)}(t) = \tilde{m}_{l}(d)  + \tilde{D}_{l}(d) x_{l}^{(B)}(t),$$
where $\tilde{m}_{l}(d) $ and $\tilde{D}_{l}(d)$ are the estimated seasonal cycles in mean and marginal variance, and $x_{l}^{(B)}(t)$ is assumed to be stationary in time. The mean seasonal cycle and the seasonal cycle of marginal variance are estimated as described in~\cite{leeds}, as also in the preceding section for the observational data. We retain $x_{l}^{(B)}(t)$, the de-seasonalized component.

Temperatures in the transient runs, on the other hand, will in general have an evolving mean in addition to nonstationarity due to seasonality. Write $y_{l,r}^{(s)}(t)$ for the temperature in the $r$'th realization of the transient scenario $s$ at time $t$ and location $l$, and assume the representation
$$ y_{l,r}^{(s)}(t) = \tilde{m}_{l}(d)  + \tilde{\mu}_{l}^{(s)}(t) + \tilde{D}_{l}(d) x_{l,r}^{(s)}(t),$$
where $\tilde{\mu}_{l}^{(s)}(t)$ represents an estimate of the evolving mean under scenario $s$ (possibly including changes in the mean seasonal cycle) and $x_{l,r}^{(s)}(t)$ is assumed to be some mean zero, but nonstationary, process. For simplicity we assume that the seasonal cycles of marginal variability do not evolve in time (overall marginal variability is still allowed to change in $x_{l,r}^{(s)}(t)$, but such changes are assumed the same for each season).  While the change in mean is needed for the simulation~\eqref{eq:simulation} (see Seciton A1 for details on its estimation), we would like to work with mean zero processes to estimate the changes in covariance structure. We therefore first remove from each transient realization the scenario average:
$$ \check{y}_{l,r}^{(s)}(t)  =  y_{l,r}^{(s)}(t)  - \frac{1}{R_s}\sum_{k=1}^{R_s}  y_{l,k}^{(s)}(t),$$
where $R_s$ is the number of realizations in the ensemble under scenario $s$. The resulting contrasts, $\check{y}_{l,r}^{(s)}(d),$ have mean zero, but still exhibit seasonal cycles in marginal variance. We thus retain the de-seasonalized contrasts
\begin{equation}
\label{eq:contrasts}
q^{(s)}_{l,r}(t) = \frac{\check{y}_{l,r}^{(s)}(t)}{\tilde{D}_{l}(d)}.
\end{equation}
While we view each run, $y_{l,r}^{(s)}(t)$, as independent, the contrasts, $q^{(s)}_{l,r}(t)$, are of course not independent across realizations within a given scenario. 

We assume that the de-seasonalized component of the preindustrial run, $x^{(B)}_{l}(t)$, has unknown marginal spectral density $\tilde{a}^{(B)}_{l}(\omega)$, and the de-seasonalized component of the transient runs, $x^{(s)}_{l,r}(t)$, has unknown evolutionary spectrum $\tilde{a}^{(s)}_{l}(t,\omega)$. While $x_{l}^{(B)}(t)$ and $x_{l,r}^{(s)}(t)$ will not be equal in distribution to the (de-seasonalized) real-world temperatures, past or future, we assume that the true changes in evolutionary spectra under scenario $s$ are equal to those of the GCM, so the de-seasonalized GCM and observed temperatures satisfy~\eqref{eq:assumptions} and, in particular,
$$ \rho^{(s,B)}_l(t,\omega) \equiv \frac{a_l^{(s)}(t,\omega)}{a_l^{(B)}(\omega)} = \frac{\tilde{a}_l^{(s)}(t,\omega)}{\tilde{a}_l^{(B)}(\omega)} .$$
In the following section, we discuss modeling and estimating these changes in evolutionary spectra.

\section{GCM projected changes in temporal covariance}
We describe a methodology for modeling and estimating the changes in covariance structure in a GCM as a function of a [CO$_2$] scenario. Our goal is not only to describe the changes in covariance in scenarios within our ensemble, but also to provide an emulator for the GCM changes in covariance for scenarios for which we have no runs. To the extent that the model we propose describes the GCM changes across the range of [CO$_2$] scenarios in our ensemble, the resulting emulator may be expected to provide good predictions of the GCM changes, at least for scenarios in some sense within the range spanned by our ensemble.

\subsection{A model for GCM changes in temporal covariance}
An important insight stated in~\cite{castruccio} is that changes in the distribution of temperatures in transient GCM runs under a [CO$_2$] forcing scenario should be describable in terms of the past trajectory of [CO$_2$]. More specifically, writing [CO$_2$]$(t)$ for the CO$_2$ concentration at time $t$, the distribution of temperature at time $t$ is determined by the function 
$$f(t') = \text{[CO$_2$]}(t-t'), \text{ for } t'>0,$$
where $f$ does not depend on $t$, so one does not need a different emulator for every time $t$. Providing useful statistical emulators for changes in the distribution of temperatures in transient GCM runs then depends on our ability to find useful functionals of the past [CO$_2$] trajectory that help explain those changes.

One potentially useful summary of the past trajectory of [CO$_2$] for a given scenario is in fact the change in regional mean temperature relative to the preindustrial climate.  We denote this change as $\bar{\Delta}^{(s,B)}_{S}(t)$ for region $S$. In this work, we have subdivided the T31 grid into the same 47 regions as in~\cite{castruccio}, chosen to be relatively homogeneous but still large enough to substantially reduce inter-annual variations. We estimate $\bar{\Delta}^{(s,B)}_{S}(t)$ in each region using a modification of the mean emulator described there (see Section A1). 

While the changes in regional mean temperature are themselves useful summaries of the past trajectory of [CO$_2$], it need not be true that temporal covariance structures will be the same if  $\bar{\Delta}^{(s,B)}_{S}(t) = \bar{\Delta}^{(s',B)}_{S}(t')$ for two scenarios $s$ and $s'$ at two different time points $t$ and $t'$. In particular, the rate of change of the evolution of regional mean temperatures ($\partial_t \bar{\Delta}^{(s,B)}_{S}(t)$ for scenario $s$) may capture some additional aspect of the changing climate that is also relevant for explaining changes in covariances.

We have indeed found that the following model usefully describes the changes in temporal covariances in scenarios in our ensemble:
\begin{equation}
\label{eq:spModel}
\log  \rho^{(s,B)}_l(t,\omega)= \delta_{l0}(\omega) \bar{\Delta}^{(s,B)}_{S}(t) + \delta_{l1}(\omega) \partial_t \bar{\Delta}^{(s,B)}_{S}(t).
\end{equation}
In the case that $\delta_{l0}(\omega)$ and $\delta_{l1}(\omega)$ are constant functions, for example,~\eqref{eq:spModel} describes a uniformly modulated process. More generally, $\delta_{l0}(\omega)$ and $\delta_{l1}(\omega)$ describe the patterns of changes in variability across frequencies associated with changes in regional mean temperature and the rate of change of regional mean temperature, respectively. 

Since each $\delta_{li}(\omega)$ is not scenario-dependent, model~\eqref{eq:spModel} can be thought of as an emulator for the GCM changes in covariance structure. That is, given an emulator for the regional mean temperature changes in the scenario of interest,~\eqref{eq:spModel} provides a prediction of the GCM changes in covariance structure under that scenario. In Section 5, we will discuss how well this model describes changes across the scenarios in our ensemble. Note that a model like~\eqref{eq:spModel} is unlikely to hold generically for all [CO$_2$] scenarios. In particular, such a model would be unlikely to fully capture the changes in variability in scenarios where [CO$_2$] changes instantaneously; such scenarios are typically not considered realistic. Additionally, since the changes in covariance in~\eqref{eq:spModel} depend on the [CO$_2$] trajectory through the corresponding \textit{changes} in mean, this model will not fully capture changes in variability that depend on \textit{absolute} temperatures through, for example, phase changes between ice and water (see Figure~\ref{fig:deviance} and discussion in Section 5.2). The model will also not fully capture GCM behavior if that behavior involved abrupt changes in the distribution of temperatures even under relatively smooth forcing scenarios (nonlinear responses to forcing); however, the model we study does not exhibit such behavior over the range of [CO$_2$] scenarios we study. We have found that the model is a useful description of changes in variability in scenarios like those in the ensemble we use here, where [CO$_2$] changes slowly and relatively smoothly over time and in locations not involving changing ice margins over the course of the scenario.

\subsubsection{Estimating $\delta_{li}(\omega)$}
To estimate the functions $\delta_{l0}(\omega)$ and  $\delta_{l1}(\omega)$, we adopt the intuitive approach for likelihoods for processes with evolutionary spectra where the usual periodogram in the Whittle likelihood is replaced with local periodograms over smaller blocks of time \citep{dahlhaus1}. While we view the local Whittle likelihood approach as most suitable in our setting, several other alternative methods for estimating evolutionary spectra have been proposed. \cite{neumann} use a wavelet basis expansion; \cite{ombao2002} use smooth, localized complex exponential basis functions; \cite{dahlhaus2} proposed another likelihood approximation that replaces the local periodogram in the earlier work with the so-called pre-periodogram introduced by~\cite{neumann}; \cite{guinness} provided an alternative generalization of the Whittle likelihood that they argued, at least in the settings they studied, is more accurate than the approximations given by either~\cite{dahlhaus1} or~\cite{dahlhaus2}. However, an advantage of the local Whittle likelihood approach is that in addition to being intuitive, the corresponding score equations are computationally easier to solve in our setting when the evolutionary spectra evolve very slowly in time so the local periodograms can be taken over large blocks of time. Computation is an especially important consideration when estimating a semi-parametric model such as~\eqref{eq:spModel}. Furthermore, the results from~\cite{guinness} suggest that the local Whittle likelihood approach may yield point estimates that are close to optimal even when the likelihood approximation itself is inaccurate, whereas for example they demonstrated that the approach based on the pre-periodogram can give unstable estimates.

In this work, we interpret the local Whittle likelihood approach as follows. We divide each contrast time series, $q_{l,r}^{(s)}(t)$, defined in~\eqref{eq:contrasts} and of length $N_s$, into blocks of length $M$ (for simplicity take $M$ to be a common factor of each $N_s$). In our setting, we take $M = 10$ years, but since temperature variability changes very slowly over time in the scenarios we analyze, the conclusions are not very sensitive to the choice of $M$; the results are essentially the same taking $M = 5$ years or $M = 30$ years, for example. Upon choosing $M$, then for the time block, location, realization, and scenario indexed by $b$, $l$, $r$, and $s$, respectively, define the local periodogram of the contrast time series at frequencies $\omega_j = 0,2\pi/M,...,2\pi$ as,
\begin{equation}
\label{eq:localPer}
I^{(s)}_{b,l,r}(\omega_j) = \frac{1}{2\pi M} \left|\sum_{t = 1}^{M} q_{l,r}^{(s)}(t+M(b-1)) e^{-i t \omega_j} \right|^2.
\end{equation}
It is straightforward to show that the Whittle likelihood for each time block, location, and scenario depends on each $I^{(s)}_{b,l,r}(\omega_j) $ only through the average across realizations,
\begin{equation}
\label{eq:avgPer}
\bar{I}^{(s)}_{b,l}(\omega_j) = \frac{1}{R_s} \sum_{r=1}^{R_s} I^{(s)}_{b,l,r}(\omega_j).
\end{equation}
Likewise, for the de-seasonalized preindustrial run, define its periodogram as
\begin{equation}
\label{eq:baselinePer}
I^{(B)}_{l}(\omega_j) = \frac{1}{2\pi N_B} \left|\sum_{t = 1}^{N_B} x_{l}^{(B)}(t) e^{-i t \omega_j} \right|^2.
\end{equation}
In our setting, since $M<N_B$, $I^{(s)}_{b,l,r}(\omega_j)$ is defined on a coarser frequency scale than is $I^{(B)}_{l}(\omega_j)$, so for the purposes of estimating changes in spectra, it may be natural to aggregate the baseline periodogram to the coarser scale; that is, write
\begin{equation}
\label{eq:avgBaselinePer}
\bar{I}^{(B)}_{l}(\omega_j) = \frac{M}{N_B} \sum_{k: \, -\frac{N_B}{2M} \leq k < \frac{N_B}{2M}} I^{(B)}_{l}\left(\frac{2\pi (j+k)}{N_B}\right).
\end{equation}

An approximate likelihood under model~\eqref{eq:spModel}, marginally at each location $l$, may then be written as the sum of the local Whittle likelihoods of the transient runs and the Whittle likelihood corresponding to the aggregated periodogram of the baseline run (under the usual approximation that the periodogram ordinates are independent at distinct Fourier frequencies):
\begin{align}
\label{eq:lik}
\mathcal{L}_{l}(\theta) =  -\frac{1}{2} \sum_{s,b,j} & \left\{(R_s-1)(\log \tilde{a}^{(B)}_l(\omega_j) + \bar{\Delta}^{(s,B)}_b \delta_{l0}(\omega_j) + \partial_t\bar{\Delta}^{(s,B)}_b \delta_{l1}(\omega_j) )\right. \\
 & + \left. R_s \bar{I}^{(s)}_{b,l}(\omega_j)e^{-(\log \tilde{a}^{(B)}_l(\omega_j) + \bar{\Delta}^{(s,B)}_b \delta_{l0}(\omega_j) +  \partial_t\bar{\Delta}^{(s,B)}_b \delta_{l1}(\omega_j))} \right\} \notag\\
 & -\frac{M}{2} \sum_j \left\{ \log \tilde{a}^{(B)}_l(\omega_j) + \bar{I}^{(B)}_l(\omega_j)/\tilde{a}^{(B)}_l(\omega_j)\right\} \notag.
 \end{align}
where $\theta= (\tilde{a}^{(B)},\delta_{l0},\delta_{l1})$ and where $\bar{\Delta}^{(s,B)}_b$ and $\partial_t\bar{\Delta}^{(s,B)}_b$ correspond to the values of $\bar{\Delta}^{(s,B)}(t)$ and $\partial_t\bar{\Delta}^{(s,B)}(t)$ for $t$ at the midpoint of time block $b$. Here the $(R_s-1)$ factor multiplying the log-determinant approximation takes into account that the contrasts, $q_{l,r}^{(s)}(t)$, are obtained by subtracting off the scenario average across realizations; see~\cite{castruccio2} for details.

The estimator maximizing~\eqref{eq:lik}, say 
\begin{equation}
\label{eq:mle}
\theta^*_l = \arg \max_{\theta}  \mathcal{L}_l(\theta),
\end{equation}
will yield very rough estimates of the functions $\tilde{a}^{(B)}_l(\omega)$, $\delta_{l0}(\omega)$, and $\delta_{l1}(\omega)$ because no smoothness has been enforced across frequencies. The baseline spectrum, $\tilde{a}^{(B)}_l(\omega)$, is not of particular interest to us, as this function is not required for the simulation~\eqref{eq:simulation}. On the other hand, maximizing~\eqref{eq:lik} is clearly inadequate for estimating the functions of interest, $\delta_{l0}(\omega)$ and $\delta_{l1}(\omega)$. 

A common approach for nonparametrically estimating the spectral density of a stationary process is to smooth its periodogram either by kernel methods or by penalized likelihood methods. For estimating ratios of spectra between two stationary processes,~\cite{leeds} adopted a penalized likelihood approach whereby the penalty enforced smoothness in the ratio. Here, we opt to smooth the rough estimates, $\delta^*_{l0}$ and $\delta^*_{l1}$, using kernel methods; that is, for $i=0,1$ write as the final estimate for $\delta_{li}$
\begin{equation}
\label{eq:smoothed_delta}
\hat{\delta}_{li}(\omega_j)   = \sum_k w_{k,j,i} \delta^*_{li}(\omega_{j+k}),
\end{equation}
where $w_{k,j,i}$ are weights (possibly varying with $j$ and $i$) satisfying $\sum_k w_{k,j,i} = 1$ for each $i$ and $j$. In practice, we use weights corresponding to a kernel with a variable bandwidth that is allowed to decrease at lower frequencies. The reason for the variable bandwidth is that in the GCM runs we have analyzed, we have observed that the log ratio of spectra are typically less smooth at very low frequencies compared to higher frequencies. For details on the form of the weights and the bandwidth selection procedure we use to choose them, see Section A2. While the penalized likelihood approach described in~\cite{leeds} may be adapted for this setting, we view the kernel smoothing approach as more straightforward, especially when allowing for variable bandwidths, and have found that the approaches yield similar estimates when the bandwidth of the kernel is constant.

Approximate pointwise standard errors for each $\hat{\delta}_{li}(\omega_j)$, and the corresponding estimate of $\log  \rho^{(s,B)}_l(t,\omega_j)$ may also be computed; these are described in Section A3. Having estimated our model, we need to compute the observation-driven simulations. Computing~\eqref{eq:simulation} efficiently is important; this is described in Section A4.

\section{Results}
We estimate the model described in Section 4 using our ensemble of CCSM3 runs described in Section 3. In this section, we describe insights into the climate system that our estimated model provides and illustrate how this information is used in our proposed simulation of temperatures, described in Section 2. In the simulation, we use temperatures from NCEP-DOE Climate Forecast System Reanalysis~\citep{Saha} as a surrogate for observational data. The reanalysis is run at T62 resolution (about a $1.875^\circ$ grid), which we regrid to the T31 resolution on which the GCM was run using an area-conserving remapping scheme. We also investigate the success of our proposed model in describing the changes in covariances observed in our GCM ensemble and the quality of our model when used as an emulator.

\subsection{Model changes in variability}
\begin{figure}[!t]
\begin{center}
\includegraphics[scale = 0.3,trim =2.75cm  0cm 0cm 0cm]{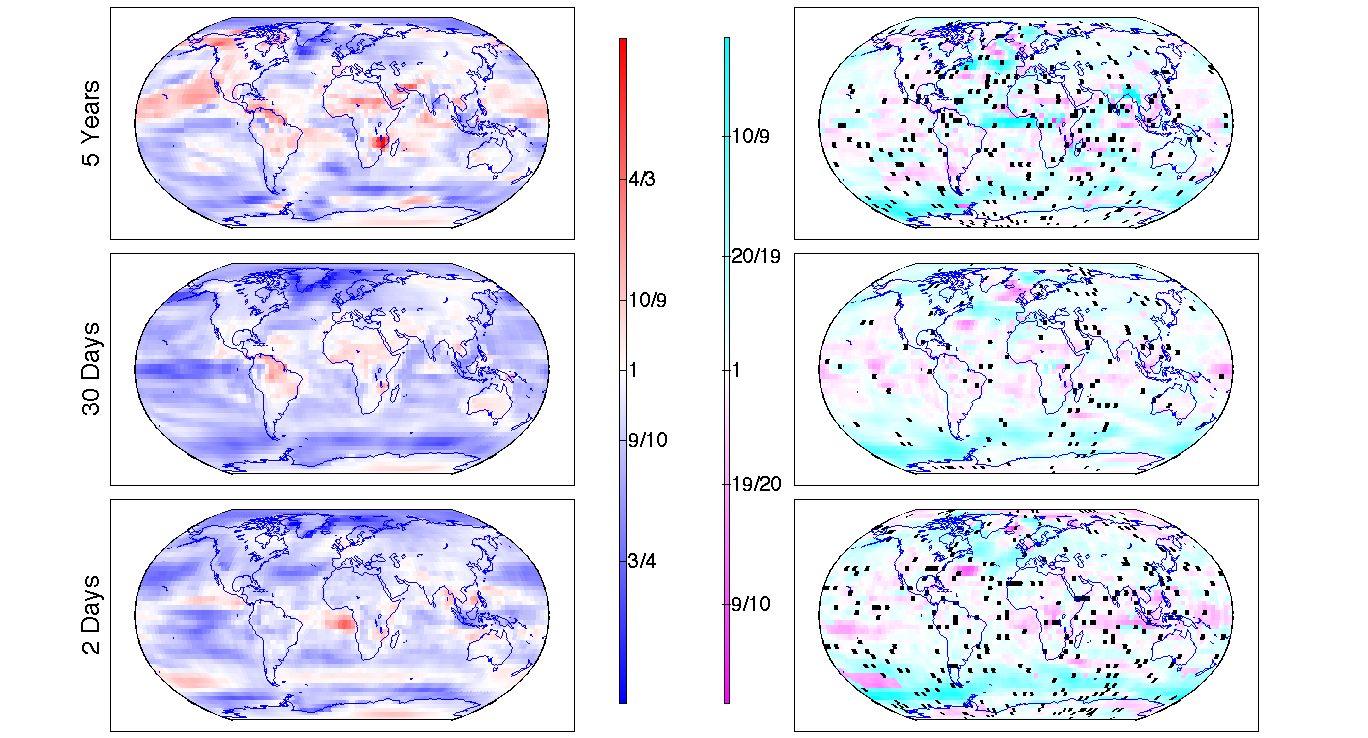}
\caption{Left, estimates of changes in marginal spectra, at three frequencies, for the low scenario at model year 2100 compared to the preindustrial climate (i.e., $\rho^{(L,B)}_l(t,\omega)$ at year 2100 and the specified frequencies). Red indicates an increase in variability and blue a decrease in variability. Right, estimates of $(\rho^{(H,B)}_l(t',\omega)/\rho^{(L,B)}_l(t,\omega))^{\textnormal{sign}_{l,t,\omega}}$ for $t'$ corresponding to the model time under the high scenario with the same change in regional mean temperature as at year 2100 in the low scenario, where $\textnormal{sign}_{l,t,\omega}$ is the sign of the log ratio at that location, time, and frequency. Magenta grid cells indicate smaller changes in variability under the high scenario at the same temperature, whereas the cyan grid cells indicate bigger changes in variability. (Black grid cells indicate the roughly 5\% of locations where the two estimates differ in sign, so comparing the relative magnitude of changes is not meaningful). Figure~\ref{fig:RequestedFig} repeats the left column maps for all the low, medium, and high scenarios.}
\label{fig:SRMap}
\end{center}
\end{figure}

In CCSM3, changes in variability in evolving climates can be primarily characterized using changes in mean temperature, with a smaller contribution by the rate of change of warming (corresponding to terms $\delta_{l0}$ and $\delta_{l1}$ in~\eqref{eq:spModel}, respectively). As a consequence, the projected patterns of changes in variability at a given time in a given future scenario largely correspond to the patterns observed in~\cite{leeds}: the GCM projects decreases in short timescale variability  at most locations, but increases in longer timescale variability in some regions, especially at lower latitudes (Figure~\ref{fig:SRMap}, left).  

The differences in variability between scenarios due to different rates of warming are small compared to the overall projected changes in variability, but also exhibit patterns. To illustrate, we compare changes in variability under the low scenario at year 2100 to the corresponding changes under the high scenario in the year of that scenario experiencing the same regional mean temperatures as at 2100 in the low scenario (Figure~\ref{fig:SRMap}, right); this year varies by region, ranging from  2037 to 2044. An analogous figure is given in Section A5 that shows the estimated changes in variability in each of the three scenarios at years corresponding to the same change in regional mean temperature.  In about 75\% of all locations, and especially in mid- and high-latitude ocean locations, the changes in variability under the high scenario are larger than under the low scenario. Larger changes under the high scenario than under the low scenario are an indication that variability changes are projected to be larger in a transient warming climate than in an equilibrated climate at the same temperature.

\begin{figure}[!t]
\begin{center}
\includegraphics[scale = 0.30,trim =4.5cm  0cm 0cm 0cm]{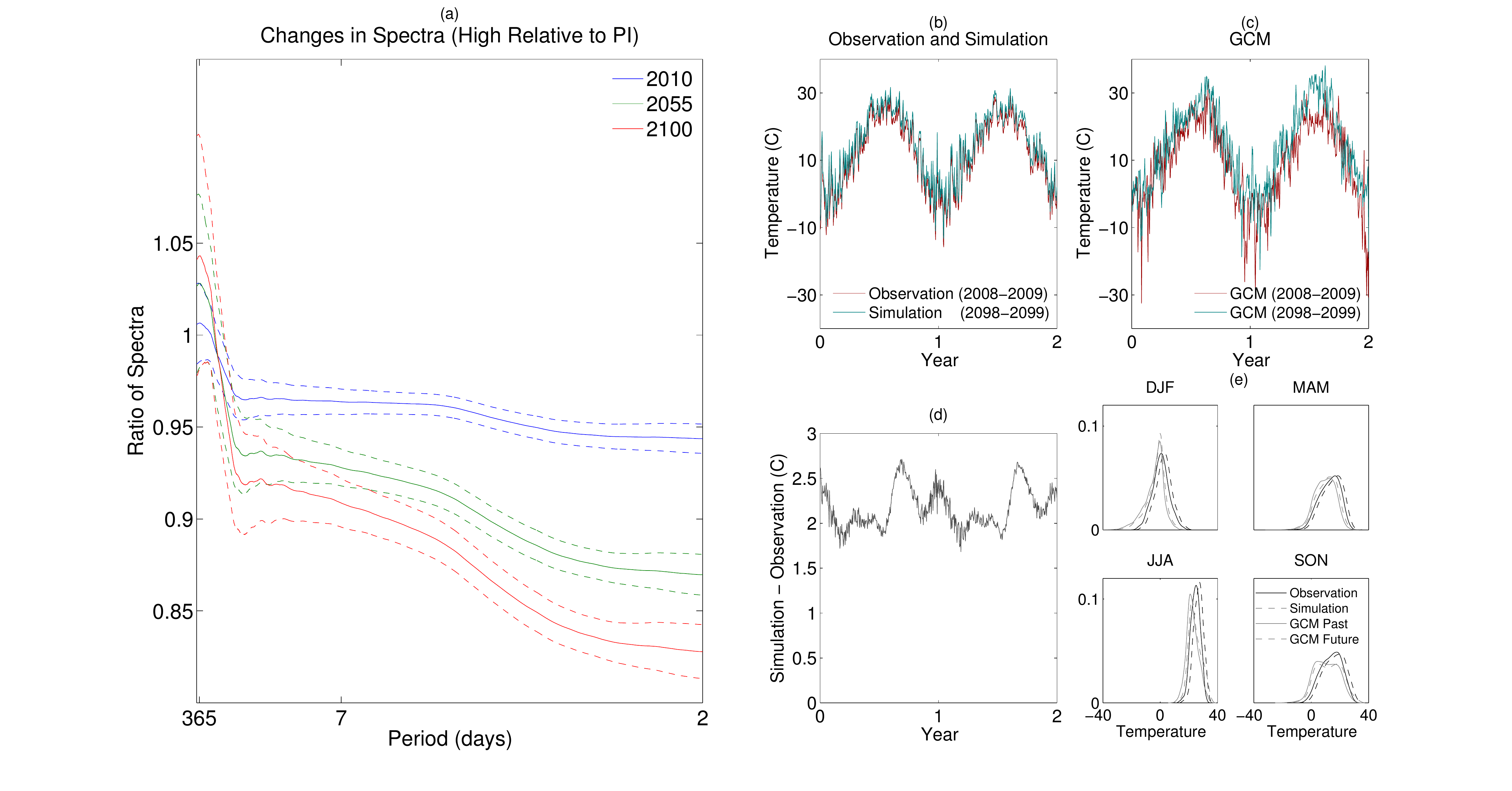}
\caption{(a), estimates of  $\rho^{(H,B)}_l(t,\omega)$ for years 2010, 2055, and 2100, at a midwestern United States grid cell (gray curves at $\pm$ two standard errors on the log scale). (b), part of the corresponding simulation computed by transforming the observational data at this grid cell via~\eqref{eq:simulation}; the simulation was computed for the whole length of the observational data, but only the last two years are displayed. (c), a run of the GCM in the years corresponding to the simulation.  (d), the difference between the observations and the simulation in panel (b). The overall shift upwards in the simulation is due to increasing mean temperature. Most of the long timescale fluctuations in the bottom panel are due to changes in the seasonal cycle: at this location, temperatures are projected to warm more in winter months than in summer months. The short timescale fluctuations on the order of ~0.1 degrees in the difference are due to changes in variability, which decreases in the future simulation. (e), marginal densities by season (labeled by corresponding months) for the observations in 2008-2009, the simulation from 2098-2099, the GCM from the same years under the historical forcing and the high future scenario scenario.}

\label{fig:ILSim}
\end{center}
\end{figure}

To illustrate how these changes in covariance structure are used in our proposed simulation, we simulate temperatures under the high scenario at a single grid cell in the Midwestern United States (Figure~\ref{fig:ILSim}, which shows the observations in 2009-2010, our simulation 89 years in the future, and output from one of the GCM runs in the same timeframes). Mean temperatures warm, more strongly in the winter than in the summer at this location, and temporal variability decreases overall. More specifically, variability is projected to modestly decrease at higher frequencies and slightly increase at low frequencies. At low frequencies, the projected log ratios are within two standard errors of zero, but at high frequencies are significantly smaller than zero. The extent to which such changes are important will of course depend on the impact domain of interest. 

The distribution of temperatures in the GCM differs strongly from that in the observations, with differences evident by eye in the raw time series and corresponding marginal densities. For example, the GCM has a stronger seasonal cycle than the observations and simulation, and greater variability in the winter months. See Section A5 for an additional comparisons of the space-time covariance structures of the observations and the GCM runs: typically, we have found that temperatures in nearby grid cells are more coherent in the GCM than in the observations, and that the coherences do not change much between the historical period and the end of the high scenario (Figures~\ref{fig:cohPlotIL}-\ref{fig:cohPlotS}). Our simulation procedure does not change the coherence structure of the observations. Collectively, this forms an argument for our procedure, which preserves features of the observations.

\subsection{Assessing model fit and quality of emulation}
To assess different aspects of how well our model describes the changes in covariance structure in CCSM3 in evolving climates, we show three diagnostics. First, we address in which geographic locations the model performs relatively better or worse. Second, we ask how well the statistical model performs as an emulator for a scenario on which the model has not been trained. Finally, we show the extent to which the rate of change of mean warming improves the quality of emulation over the simpler model where changes in covariance are explained solely by changes in mean temperature.

\begin{figure}[!t]
\begin{center}
\includegraphics[scale = 0.25,trim =0cm  2cm 0cm 0cm]{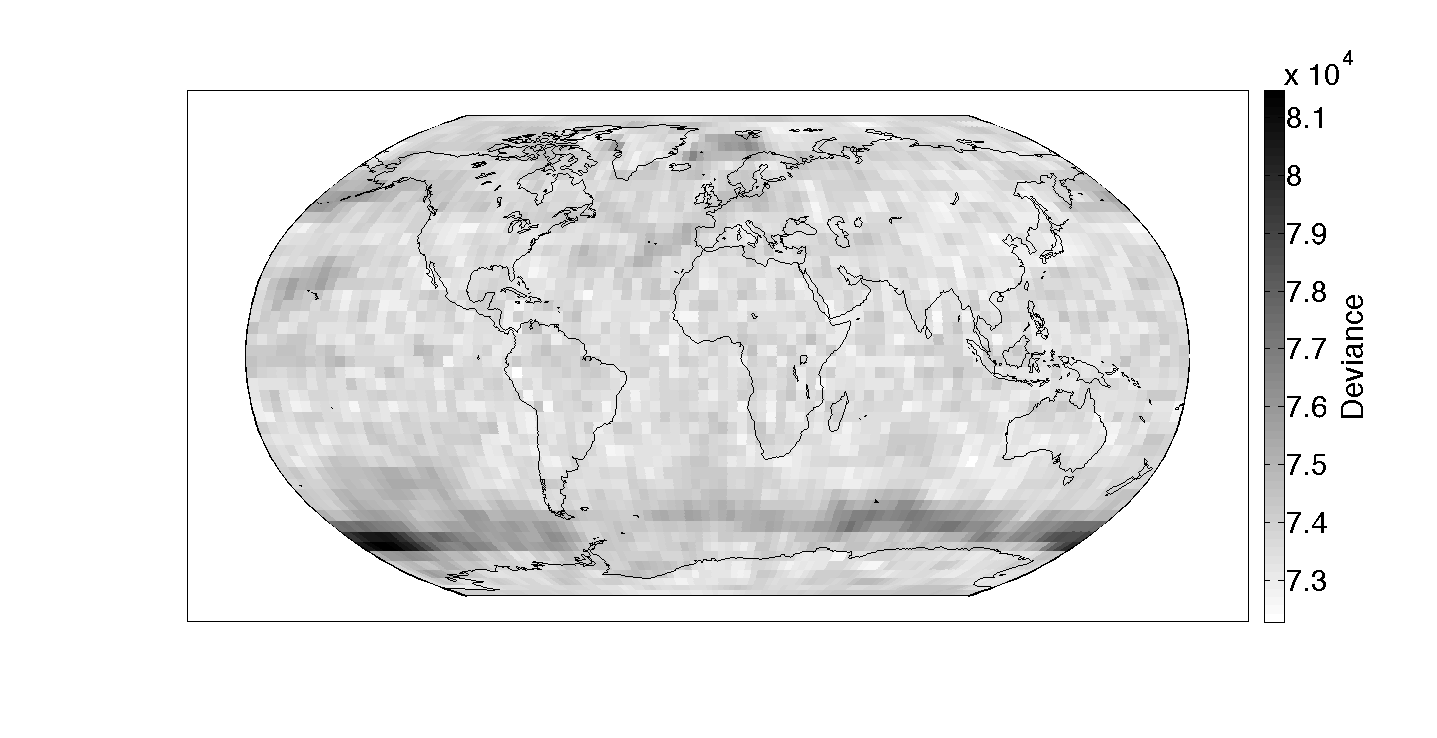}
\caption{Deviances, comparing the approximate likelihood under our estimated model to that under the saturated model where the spectrum in each time block, scenario, and frequency has its own parameter. The number of observations at each location is about 1.6 million days. The regions showing the largest deviances are those where changes in sea ice have a strong influence on variability; in such locations, our model based on changes in mean temperature cannot be expected to be a fully adequate description of changes in variability.}
\label{fig:deviance}
\end{center}
\end{figure}

To examine in which locations the model performs best and worst, we compare the deviances of our model at each location (Figure~\ref{fig:deviance}); recall that the deviance compares the likelihood under our estimated model to that under the saturated model where the value of the spectrum in each time block, scenario, and frequency is assigned its own parameter. The deviances are largest at the edge of the maximum present-day sea ice extent in the Southern Ocean, and relatively homogeneous elsewhere. The relatively poorer fit at ice margins is expected, since variability decreases substantially here as sea ice retreats, and those changes are therefore based in part on absolute temperatures.  Any statistical model based purely on changes in temperature, rather than absolute temperatures, will have difficulties capturing variability changes due to phase changes between ice and water. This result should serve as a warning against using such methods over locations, scenarios, and time periods in which the response to [CO$_2$] changes is highly nonlinear.

\begin{figure}[!t]
\begin{center} 
\includegraphics[scale = 0.365,trim =2cm  2cm 0cm 0cm]{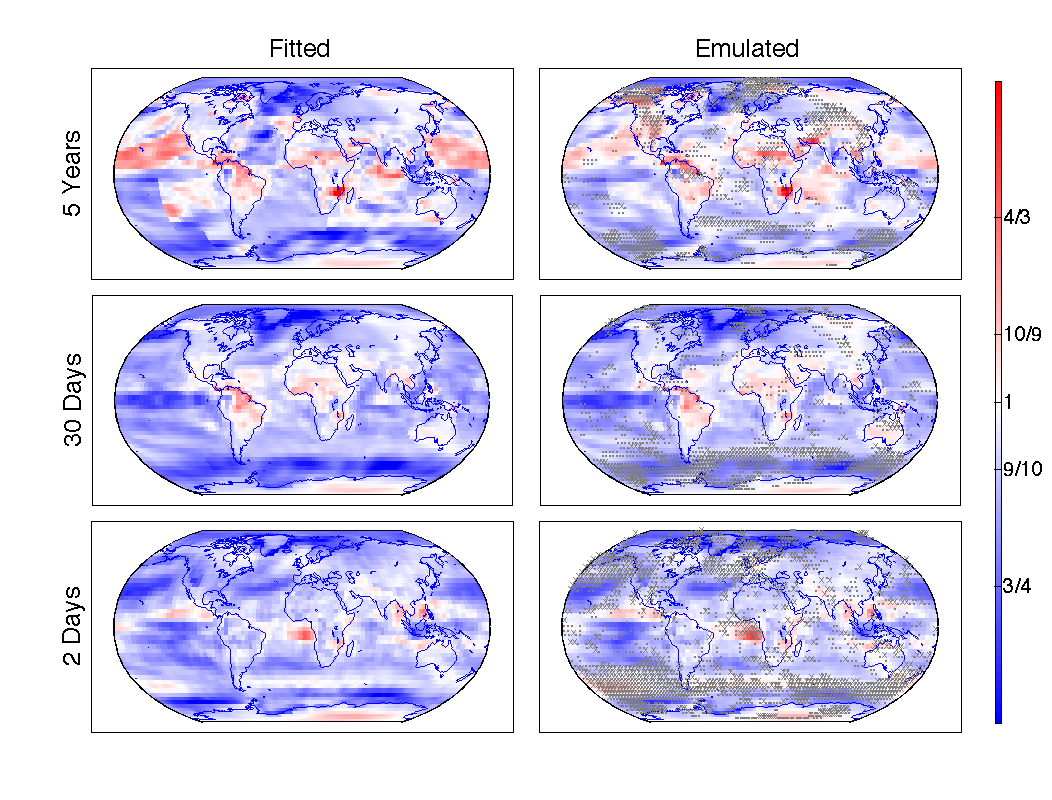}
\caption{Estimates of changes in marginal spectra, at three frequencies, for the medium scenario at model year 2100 compared to the preindustrial climate. Left, estimates use only the medium scenario realizations. Right, estimates use all but the medium scenario realizations (i.e., are estimated as an emulator).  Note that the apparent sudden change in behavior for low frequency variability in the Pacific Ocean (top left) is an artifact of the minimum bandwidth chosen for smoothing in these two adjacent regions; see Section A2 for details on bandwidth selection. Locations are marked with ``.'' (or ``x'') when the difference between the emulator and the fitted model is more than two (or three) standard errors away from zero. The patterns are similar under both schemes, with most of the differences at locations where our model would not be expected to be a good description of changes in variability (e.g., at ice margins).}
\label{fig:emMed}
\end{center}
\end{figure}

To address how well our statistical model is able to emulate GCM projected changes in variability for scenarios in some sense within the range spanned by our ensemble, we re-estimate our model using (a) all but the realizations under the medium scenario, and (b) only the realizations under the medium scenario. If the conclusions we draw from (a) match those drawn from (b), this is evidence that we have successfully emulated the changes in covariance structure under the medium scenario. We compare projected changes in marginal evolutionary spectra at year 2100 of the medium scenario, estimated under these two schemes (Figure~\ref{fig:emMed}). The estimated global patterns of changes in variability are quite similar under the two schemes. In an absolute sense, the biggest differences between the two schemes are at the lowest frequencies, but recall that since we have reason to believe that the ratios of spectra are less smooth at lower frequencies, we smooth with a smaller bandwidth at those frequencies and therefore our estimates of these changes are more uncertain. Globally, the differences between the estimates under schemes (a) and (b) are within two standard errors of zero in about 60-75\% of the grid cells, depending on the frequency of interest. (Over land, the differences are within two standard errors in about 70-80\% of the grid cells.)  The locations where there are significant differences between the two schemes are, unsurprisingly, often those where we have argued that the model should have trouble, such as in the Southern Ocean. In these locations, the emulator usually underestimates variability changes.

\begin{figure}[!t]
\begin{center}
\includegraphics[scale = 0.25,trim =0cm  2cm 0cm 0cm]{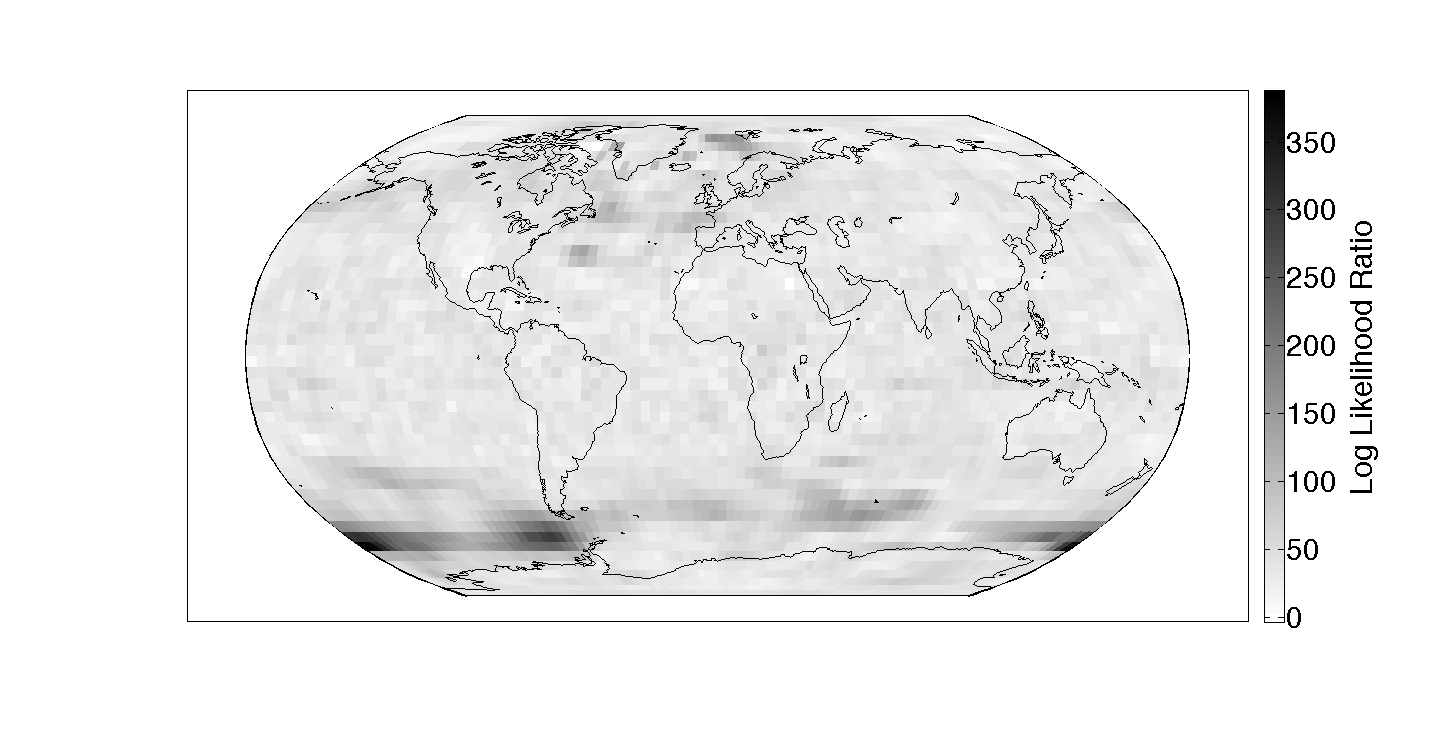}
\caption{Log likelihood ratios comparing our proposed model to the reduced model taking $\delta_{l1}(\omega)=0$. The models were estimated using all but the medium scenario realizations and the likelihoods were computed with just the medium scenario realizations, so the likelihood ratio is a comparison of the quality of emulation (no adjustment for model complexity is needed). In all but three locations on the globe, the likelihood under the full model is larger, which is an indication that for the purposes of emulation it is useful to include $\delta_{l1}(\omega)$ in the model.}
\label{fig:LRMedEm}
\end{center}
\end{figure}

As discussed above, a feature of our model is that changes in variability depend not only on the change in regional mean temperature but also on the rate at which those changes occur. One might ask whether the simpler model that omits the second term (i.e., $\delta_{l1}(\omega)=0$) is just as good at emulating changes in variability. Figure~\ref{fig:LRMedEm} displays the predictive log likelihood ratio comparing the simpler model to our proposed model; by predictive log likelihood, we mean that the models are estimated as emulators, excluding the medium scenario realizations, and the likelihoods were evaluated for the medium scenario realizations (as such, no adjustment for model complexity is necessary). At all but three out of the 4,608 locations, the full model has a larger predictive likelihood than the simpler model, which indicates that for the purposes of emulation it is useful to allow for a nonzero term involving the rate of change change of warming.

\section{Discussion}
In this work, we describe a method for transforming observed temperatures to produce simulations of future temperatures when the climate is in a transient state, based on the projected changes in means and temporal covariances in GCM output. We believe this approach should yield more realistic simulations of future climate than do either GCM runs or simulations based on modifying GCM runs.  Any observation-driven procedure is, however, of course limited by the observational record: for example, as described, our procedure provides exactly one simulation of future temperatures equal in length to the observational record. As also suggested by~\cite{leeds}, longer simulations could be produced either by recycling the observations entirely or by resampling them to generate new pseudo-observations.

An important feature of the procedure we describe is that our simulations preserve many features of the observational record not accounted for explicitly in the procedure. This paper is concerned with changes in only the temporal covariance structure of temperatures in transient climates. In the methodology described here, the simulation therefore preserves, for example, the spatial coherence spectra of the observations. While projected changes in spatial coherences in the model we study appear to be small (Section A5), changes in spatial coherences may also be important for societal impacts. We leave for future research the possibility of extending the methodology to account for such changes. This extension would be challenging and interesting, in part because temperatures are nonstationary in space with abrupt, local changes due to geographic effects.

Another challenging and interesting extension of our methods would be to jointly simulate future temperature and precipitation. Our work has focused on simulating temperatures, but potential changes in precipitation are also important for societal impacts. While there have been some model-driven proposals for jointly simulating temperature and precipitation~\citep{piani,vrac}, to our knowledge most approaches (both model- and observation-driven) proceed by simulating the two quantities separately. Versions of the simple Delta method can and are used for monthly precipitation, with the assumption that the GCM captures multiplicative (rather than additive) changes in rainfall amount (see, e.g.,~\cite{teutschbein} and references therein for a review of common precipitation simulation methods). A simple Delta method cannot, however, capture the changes on the timescale of individual rainfall events, whose intensity changes differently than that of time-averaged rainfall, with projections of less frequent but more intense storms (e.g.,  \cite{trenberth2011}). Our approach, based on spectral methods, is likely also inadequate for characterizing changes in variability of daily precipitation, because daily precipitation often takes the value zero. More sophisticated observation-driven simulations methods for precipitation remains an area of research, as does the joint simulation of temperature and precipitation. In the context of this paper, where changes to the correlation structure of temperatures are detectable but not very large, we find it unlikely that separately simulating temperature and precipitation (as in common practice) will result in large changes to their bivariate dependence structure.

The characterization of variability changes in transient climates is itself an issue of scientific interest. One of our motivations for developing the methods described here is to enable studying and ultimately comparing different GCM projections of changes in temperature variability. Most publicly available GCM runs (including most runs mandated by the IPCC) describe plausible future climates, which are necessarily in transient states. The statistical model we develop uses the GCM change in regional mean temperature and its rate of change to describe the GCM projected changes in covariance structure; in the GCM runs we study, these factors effectively summarize the projected changes in covariance. While we have investigated changes in variability in only one GCM, at relatively coarse resolution, we hope that our methods are applicable across GCMs, and may aid in carrying out a comparison across different GCMs in a coherent and interpretable way.

\section*{Acknowledgements}
We thank William Leeds and Joseph Guinness for helpful discussions and for providing code related to their manuscripts. Support for this work was provided by STATMOS, the Research Network for Statistical Methods for Atmospheric and Oceanic Sciences (NSF-DMS awards 1106862, 1106974, and 1107046), and RDCEP, the University of Chicago Center for Robust Decision Making in Climate and Energy Policy (NSF grant SES-0951576).

\bibliographystyle{plainnat}
\bibliography{bibNotes}

\renewcommand{\thesection}{A\arabic{section}}   
\renewcommand{\thetable}{A\arabic{table}}   
\renewcommand{\thefigure}{A\arabic{figure}}
\renewcommand{\theequation}{A\arabic{equation}}
\setcounter{section}{0}
\setcounter{figure}{0}
\setcounter{equation}{0}
\setcounter{table}{0}

\section{Estimating changes in regional and local mean temperature}
\begin{figure}
\begin{center}
\includegraphics[scale = 0.25,trim =0cm  2cm 0cm 0cm]{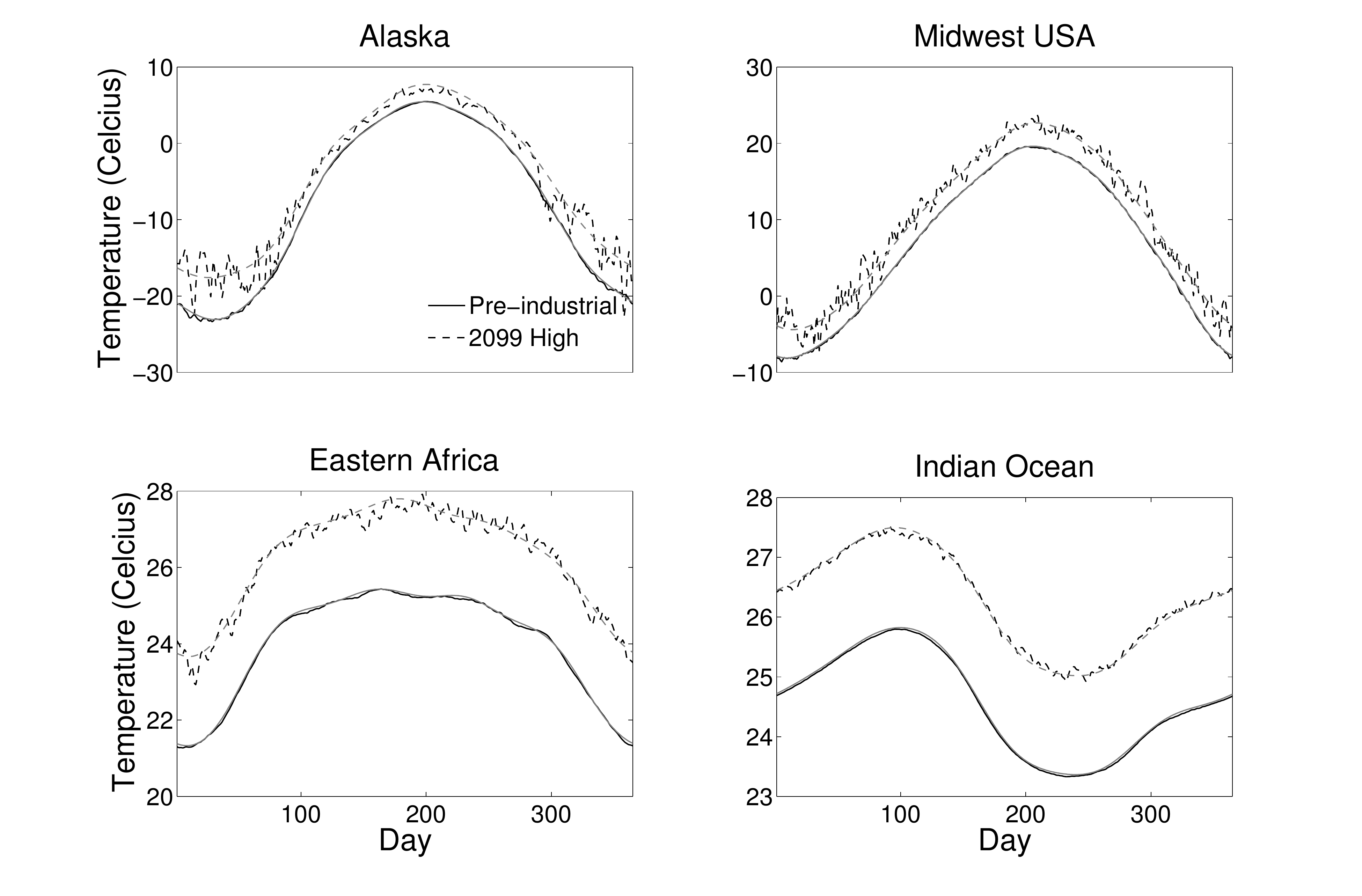}
\caption{Examples of emulated (gray) and empirical (black) mean seasonal cycles in four regions compared between the preindustrial climate (solid) and the climate under the high scenario in the year 2099 (dashed). The emulations are fitted according to~\eqref{eq:MeanEmulator}. The empirical estimates are obtained by averaging across grid cells in the region of interest and across time (for the equilibrated preindustrial run) or across realizations (for the high scenario realizations). The empirical estimates of the 2099 seasonal cycles (red) are noisy because they represent averages across only eight realizations of the regional mean temperature.}
\label{fig:seasonalCycles}
\end{center}
\end{figure}

The model for changes in covariance,~\eqref{eq:spModel}, requires an estimate of the changes in regional mean temperature, $\bar{\Delta}^{(s,B)}_{S}(t)$, for scenario $s$. We also need changes in local mean temperature to compute the simulation~\eqref{eq:simulation}. We estimate these using a modification of the mean emulator described in~\cite{castruccio}. Write $\bar{\mu}_{S}^{(s)}(t)$ for the regional mean temperature at time $t$ under scenario $s$ in region $S$. We assume that
\begin{align}
\label{eq:MeanEmulator}
\bar{\mu}_{S}^{(s)}(t) & = &  \beta_{0,S} + \beta_{1,S}C^{(s)}(t)+ \sum_{k=1}^K\left\{\gamma_{k,S} \cos\left(\frac{2\pi t k}{365}\right) + \zeta_{k,S} \sin\left(\frac{2\pi t k}{365}\right) \right\} \\
 &  & +  \sum_{k=1}^K\left\{\gamma'_{k,S} C^{(s)}(t) \cos\left(\frac{2\pi t k}{365}\right) + \zeta'_{k,S} C^{(s)}(t)  \sin\left(\frac{2\pi t k}{365}\right) \right\} \notag
\end{align}
where 
$$C^{(s)}(t) = \sum_{m=0}^{\infty} (1-\phi_S) \phi_S^{m} \log\left(\frac{[\text{CO}_2]^{(s)}(t-m)}{[\text{CO}_2]^{(\text{B})}}\right)$$
 for $\phi_S\in[0,1)$, and $[\text{CO}_2]^{(s)}(t)$ and $[\text{CO}_2]^{(\text{B})}$ are the CO$_2$ concentrations under scenario $s$ at time $t$ and under preindustrial conditions, respectively. For the harmonic terms, we take $K=6$. This mean emulator differs from that described in~\cite{castruccio} in essentially two ways. First, here we exclude one effect in theirs that was meant to distinguish short term and long term effects of changes in [CO$_2$]. For smooth, monotonic scenarios like those in our ensemble, it is difficult to distinguish these two effects. Second, whereas their paper was concerned with emulating annual average temperatures, here we are interested in daily temperatures, so we need terms that account for the (possibly changing) mean seasonal cycle; note that the GCM runs use a year of exactly 365 days, hence our representation of the seasonal cycle in terms of harmonics of $1/365$. In most regions, any changes to the mean seasonal cycle are small besides an overall increase in mean, although in regions with strong seasonal cycles, the mean seasonal cycle tends to be damped more in winter months than summer months (Figure~\ref{fig:seasonalCycles}).
 
For the change in regional mean temperature used as an input in~\eqref{eq:spModel}, we define $\bar{\Delta}^{(s,B)}_{S}(t) = \beta_{1,S}C^{(s)}(t)$, which is, under the above model, the change in regional mean temperature from the preindustrial climate excluding changes in the seasonal cycle. For the changes in local mean temperature used as an input in the simulation~\eqref{eq:simulation}, we assume that the local means are related to the regional means through a regional pattern scaling relationship; that is, the change in local mean from the preindustrial climate is taken to be proportional to the change in regional mean from preindustrial climate:
$$\tilde{\mu}^{(s)}_{l}(t) - \tilde{\mu}^{(B)}_{l}  = \lambda_l(\bar{\mu}_{S}^{(s)}(t) - \bar{\mu}_{S}^{(B)}).$$
This is the approach advocated by~\cite{castruccio} for grid cell-level mean emulation.

\section{Choice of variable bandwidth for estimating $\delta_{li}(\omega)$}
\begin{figure}
\label{fig:bandwidths}
\begin{center}
\includegraphics[scale = 0.25,trim =0cm  0cm 0cm 0cm]{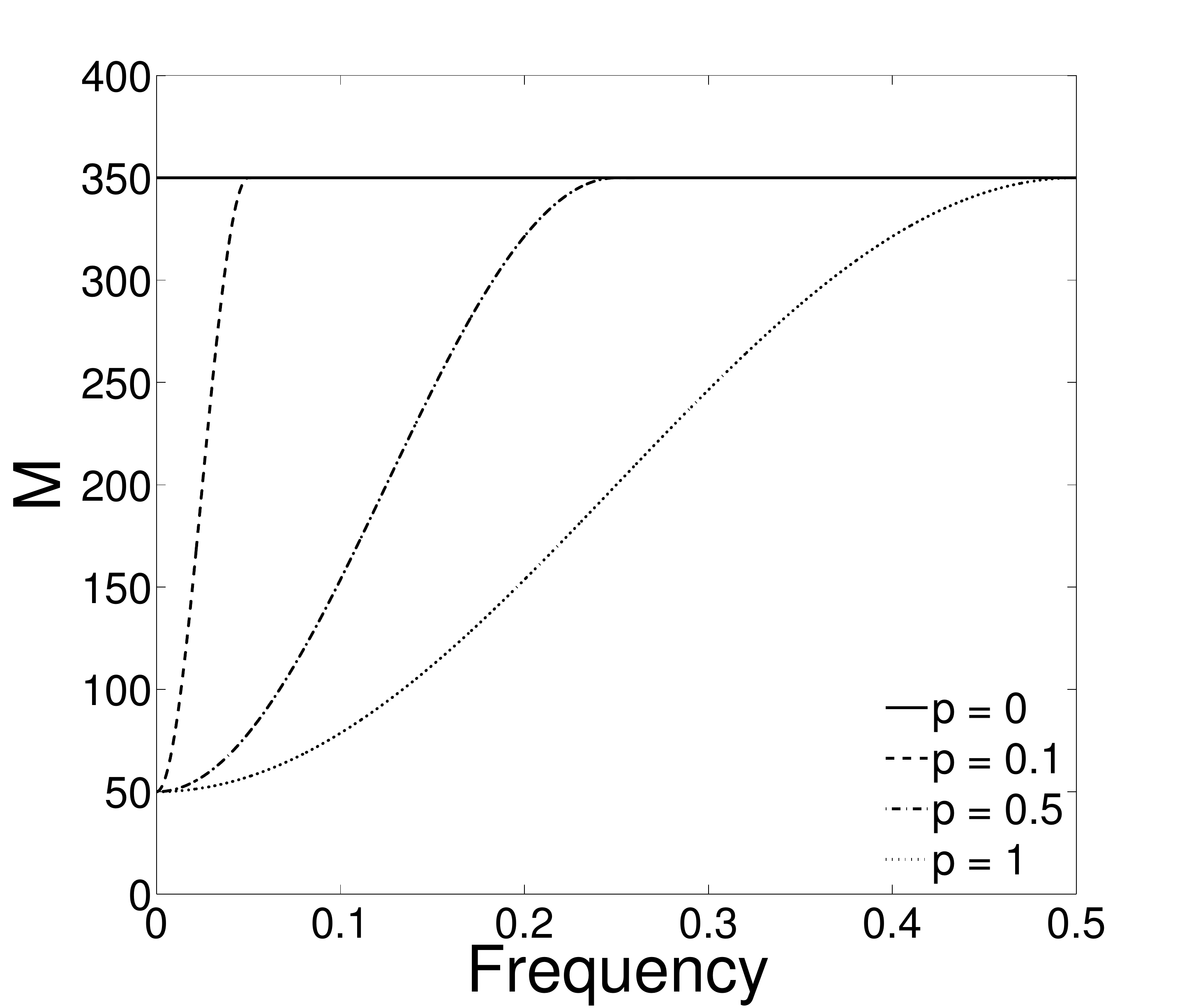}
\caption{Examples of the function $M_{j,i}$ controlling the bandwidth of the kernel estimator as a function of the frequency being estimated. In each of these curves $m_{0,i} = 50$ (controlling the minimum bandwidth) and  $m_{1,i} = 350$ (controlling the maximum bandwidth), and the value of $p_i$ (controlling the transition between the minimum and maximum bandwidths) is noted in the figure. The form of $M_{j,i}$ is such that the bandwidth of the kernel is smaller at low frequencies.}
\end{center}
\end{figure}

Here we describe the choice of weights, $w_{k,j,i}$, used in the estimates of $\delta^{(s)}_{l,i}(\omega_j)$ given by~\eqref{eq:smoothed_delta}. We let $w_{k,j,i}$ correspond to a variable-bandwidth quadratic kernel,
$$
w_{k,j,l}\propto \begin{cases}
1-(\frac{k}{M_{j,i}+1})^{2}, & k\in\{-M_{j,i},...,M_{j,i}\}\\
0, & \mbox{otherwise}
\end{cases}
$$
where $M_{j,i}$ controls the bandwidth at frequency $\omega_j$ for $i=0,1$, which we allow to vary as $M_{j,i}=(m_{1,i}-m_{0,i})h_{j}(p_i)+m_{0,i}$ where $m_{1,i}$ and $m_{0,i}$ are such that $0 < m_{0,i} \leq m_{1,i}$ and
$$
h_{j}(p_i) = \begin{cases}
\frac{1}{2}(1+\cos\left(\frac{2\pi}{p_i}(\omega_{j}-\frac{p_i}{2})\right)), & 0\leq\omega_{j}<\frac{p_i}{2}\\
1 & p_i\leq\omega_{j}\leq1-p_i\\
\frac{1}{2}(1+\cos\left(\frac{2\pi}{p_i}(\omega_{j}-\frac{p_i}{2})\right)) & 1-\frac{p_i}{2}<\omega_{j}<1,
\end{cases}
$$
so $m_{1,i}$ controls the maximum bandwidth, $m_{0,i}$ the minimum bandwidth, and $p_i$ the frequency at which the bandwidths transition between $m_{0,i}$ and $m_{1,i}$ (see Figure~\ref{fig:bandwidths}).  The reason we allow the bandwidths to decrease in frequency in this constrained way is that we have found that the log ratio of spectra in our ensemble are typically less smooth at very low frequencies. The specific form of $h$ is somewhat arbitrary, but appears to work well in practice.

To implement the estimator, we need to choose the parameters $m_{1,i}$, $m_{0,i}$, and $p_i$. This is done via leave-one-out cross validation; that is, for each region the parameters are chosen to minimize the sum of the cross-validation scores for each location,
$$R_{CV}(m_{1,i}, m_{0,i}, p_i) = \sum_{j,l} \frac{(\hat{\delta}_{li}(\omega_j) - \delta^*_{li}(\omega_j))^2}{(1-w_{0,j,i})^2},$$
where $\delta^*_{li}$ is the maximizer of the approximate likelihood,~\eqref{eq:lik}, and $\hat{\delta}_{li}$ is the smoothed version with weights $w_{k,j,l}$ defined above. These parameters vary by region, but typical values of $(m_{1,i},m_{0,i},p_i)$ when estimating $\delta_{l0}$ and $\delta_{l1}$ using all runs in our ensemble are around $(350, 50, 0.4)$ for $i=0$ and $(800, 400, 0.7)$ for $i=1$ (recall that the local periodograms in~\eqref{eq:lik} are taken over ten-year blocks, so the raw estimates of $\delta_{li}$ are calculated at 1,825 frequencies). Since the parameters vary regionally, there may be sudden changes in estimates of the changes in spectra near boundaries of the region (see Figure~\ref{fig:emMed}, top right). One could consider a post-hoc spatial smoothing of either the bandwidths or the changes in spectra to avoid this, but this is not explored in this paper.

\section{Standard errors for estimates of $\delta_{li}(\omega)$}
Under~\eqref{eq:lik}, the Fisher information for $(\delta_{l0},\delta_{l1},\tilde{a}^{(B)})$ is a block diagonal matrix with blocks for each frequency, $\omega_j$, approximately (under standard approximations for periodograms) equal to 
\begin{equation}
\label{eq:information}
\mathcal{I}(\omega_j) = \left[\begin{array}{ccc}
\sum_{sb}(\bar{\Delta}_{b}^{(s,B)})^{2}R_{s}' & \sum_{sb}(\partial_{t}\bar{\Delta}_{b}^{(s,B)})\bar{\Delta}_{b}^{(s,B)}R_{s}' & \sum_{sb}\frac{\bar{\Delta}_{b}^{(s,B)}R_{s}'}{\tilde{a}^{(B)}(\omega_{j})}\\
 & \sum_{sb}(\partial_{t}\bar{\Delta}_{b}^{(s,B)})^{2}R_{s}' & \sum_{sb}\frac{\partial_{t}\bar{\Delta}_{b}^{(s,B)}R_{s}'}{\tilde{a^{(B)}}(\omega_{j})}\\
 &  & \sum_{sb}\frac{R_{s}'+M}{\tilde{a}^{(B)}(\omega_{j})^{2}},
\end{array}\right]
\end{equation}
where $R_s'=R_s-1$. We calculate standard errors under the usual approximation that the variance of the maximum likelihood estimate§ is the inverse information matrix; such an approximation can be justified by, for example, considering either the number of independent GCM runs or the number of time blocks to be large. The resulting covariance matrix for $\delta_{l0}^*(\omega_j)$ and $\delta_{l1}^*(\omega_j)$, the maximum likelihood estimators for the $\delta$ functions, is constant in frequency. Write 
\begin{eqnarray*}
\mbox{Var}\left(\begin{array}{c} \delta_{l0}^*(\omega_j) \\ \delta_{l1}^*(\omega_j) \end{array}\right) & \approx & \left[\begin{array}{cc}
\mathcal{I}(\omega_j)^{-1}_{1,1} &\mathcal{I}(\omega_j)^{-1}_{1,2} \\
 &\mathcal{I}(\omega_j)^{-1}_{2,2}\end{array}\right]\\
 & \equiv & \left[\begin{array}{cc}
V_{\delta_{l0},\delta_{l0}} & V_{\delta_{l0},\delta_{l1}} \\
 & V_{\delta_{l1},\delta_{l1}}\end{array}\right].
\end{eqnarray*}
Then the variance of the smoothed estimates, $\hat{\delta}$, defined in \eqref{eq:smoothed_delta}, at frequencies not close to 0 or $\pi$ is
\begin{eqnarray*}
\label{eq:varEst}
\mbox{Var}\left(\begin{array}{c} \hat{\delta}_{l0}(\omega_j) \\ \hat{\delta}_{l1}(\omega_j) \end{array}\right)
& \equiv & \left[\begin{array}{cc}
V_{\hat{\delta}_{l0},\hat{\delta}_{l0}} & V_{\hat{\delta}_{l0},\hat{\delta}_{l1}} \\
 & V_{\hat{\delta}_{l1},\hat{\delta}_{l1}}\end{array}\right] \\
& = & \left[\begin{array}{cc}
V_{\delta_{l0},\delta_{l0}}\sum_k w_{k,j,0}^2& V_{\delta_{l0},\delta_{l1}}\sum_k w_{k,j,0}w_{k,j,1} \\
 & V_{\delta_{l1},\delta_{l1}}\sum_k w_{k,j,1}^2\end{array}\right].
\end{eqnarray*}
When 0 or $\pi$ is within the local bandwidth for the frequency of interest, we make the correction to the above to account for the fact that $\delta_{li}^*(\omega_j)$ is periodic and even (here omitted for simplicity).  The variance of $\log \hat{\rho}_l^{(s,B)}(t,\omega)$, the smoothed estimate of log ratio of spectra, is then approximately
\begin{align*}
\mbox{Var} \log \hat{\rho}_l^{(s,B)}(t,\omega) \approx  (\bar{\Delta}^{(s,B)}_{S}(t))^2 V_{\hat{\delta}_{l0},\hat{\delta}_{l0}} + (\partial_t \bar{\Delta}^{(s,B)}_{S}(t))^2 V_{\hat{\delta}_{l1},\hat{\delta}_{l1}} \\ 
										+  2 \bar{\Delta}^{(s,B)}_{S}(t) \partial_t \bar{\Delta}^{(s,B)}_{S}(t) V_{\hat{\delta}_{l0},\hat{\delta}_{l1}}.
\end{align*}

\section{Computing simulations}
Computing~\eqref{eq:simulation} efficiently requires the ability to quickly compute the products $C_{N_T}\left(\sqrt{\hat{\rho}^{(s,B)}}\right)x$ and $C^{-1}_{N_T}\left(\sqrt{\hat{\rho}^{(0,B)}}\right) x$ for a vector $x$. The matrix-vector products may be written as 
\begin{align*}
\left(C_{N_T}\left(\sqrt{\hat{\rho}^{(s,B)}}\right)x\right)_t   =  \sqrt{\frac{2\pi}{N_T}} \sum_{j=0}^{N_T-1}x_{j}e^{\frac{1}{2}[\bar{\Delta}^{(s,B)}_t \hat{\delta}_{l0}(\omega_j) + \partial_t\bar{\Delta}^{(s,B)}_t \hat{\delta}{l1}(\omega_j)]}e^{i\omega_{j}t}\\
  =  \sqrt{\frac{2\pi}{N_T}} \sum_{j=0}^{N_T-1} x_{j} \sum_{p=0}^{\infty} \frac{\{\frac{1}{2}[\bar{\Delta}^{(s,B)}_t \hat{\delta}_{l0}(\omega_j) + \partial_t\bar{\Delta}^{(s,B)}_t \hat{\delta}_{l1}(\omega_j)]\}^p}{p!}e^{i\omega_{j}t} \\
  \approx \sqrt{\frac{2\pi}{N_T}} \sum_{p=0}^{P-1} \sum_{m=0}^p \frac{ (\bar{\Delta}^{(s,B)}_t)^{p-m} (\partial_t \bar{\Delta}^{(s,B)}_t)^{m}}{2^p m! (p-m)!} \sum_{j=0}^{N_T-1} \hat{\delta}_{l0}(\omega_j)^{p-m} \hat{\delta}_{l1}(\omega_j)^{m} x_j e^{i\omega_j t}
\end{align*}
where the approximation just truncates the Taylor series at $P-1$, then uses the binomial expansion and changes the order of summation. This approximation is the weighted sum of $P(P+1)/2$ inverse discrete Fourier transforms; if $P$ can be taken to be only modestly large so that the Taylor approximation is accurate, this sum can be computed efficiently. In our setting, the changes in variability projected by the GCM are small and $P$ need not be very large in order for the approximation to be quite good; we have found that taking $P=10$ is more than enough to give accurate approximations given the magnitude of the estimated changes in variability.

To compute the matrix-inverse vector products, we have found, as in~\cite{guinness}, that iterative methods work well. In order to work well, these require the ability to compute forward multiplication quickly, which we have just described, as well as a good preconditioner; since the historical time series is only mildly nonstationary, we have found that the Fourier transform scaled by the square root of the average of $\rho^{(0,B)}(t,\omega)$ over time works well as a preconditioner.

\newpage

\section{Additional supplementary figures}
\begin{figure}[!h]
\begin{center}
\includegraphics[scale = 0.33,trim =0cm  0cm 0cm 0cm]{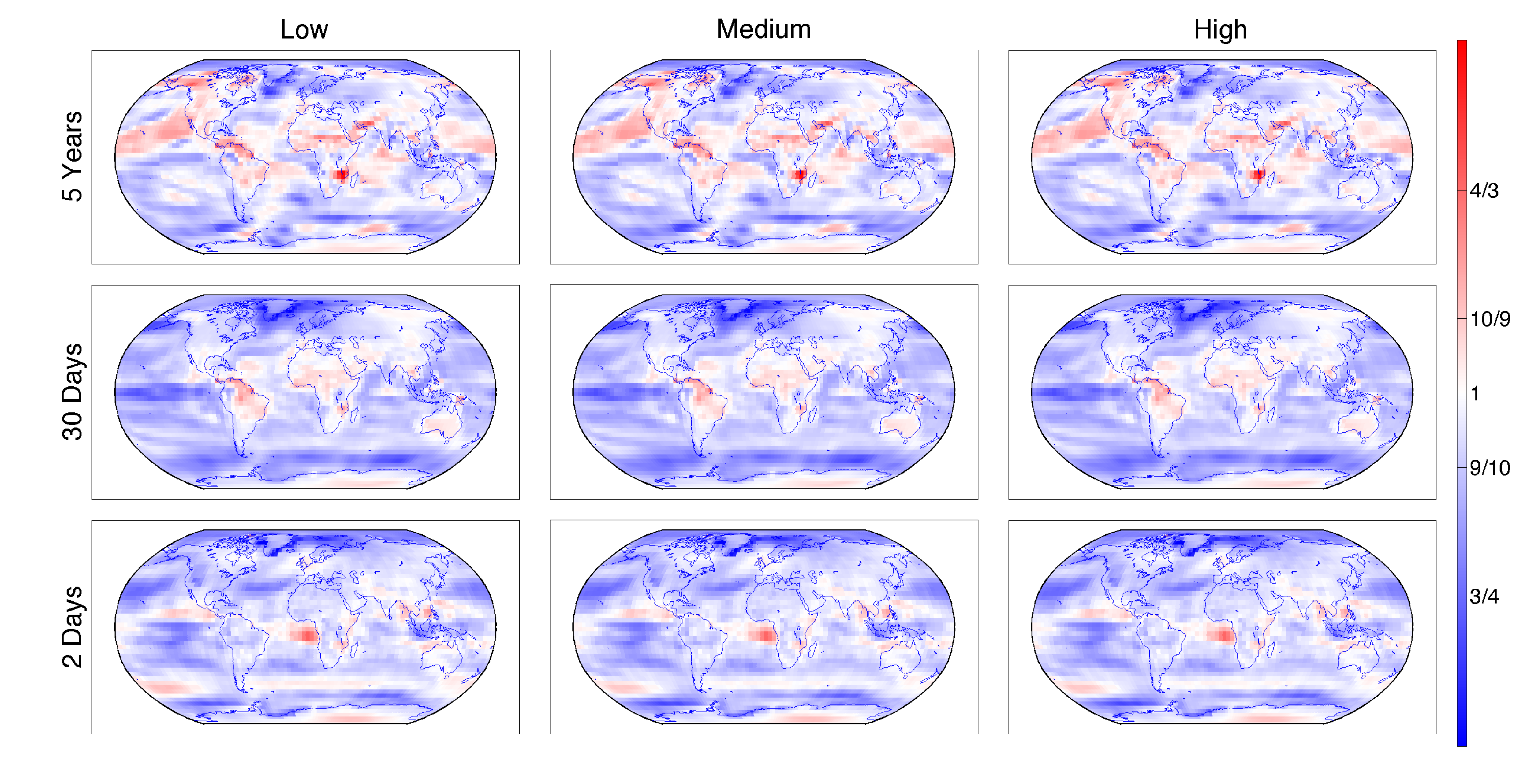}
\caption{Estimates of changes in marginal spectra, at three frequencies, at the year corresponding to that with the same regional mean temperature as in the low scenario in model year 2100 (i.e., left, $\rho^{(L,B)}_l(t,\omega)$ at year 2100; middle, $\rho^{(M,B)}_l(t,\omega)$ at years ranging from 2050-2056 depending on region; right, $\rho^{(H,B)}_l(t,\omega)$ at years ranging from 2037-2044 depending on region). Red indicates an increase in variability and blue a decrease in variability. The estimated differences between scenarios are small because the effect of the rate of change of warming on the estimated changes in variability are smaller than the effect of the overall regional mean change in temperature. A direct comparison of the estimates between the high and low scenario is shown in Figure~\ref{fig:SRMap} of the main text.}
\label{fig:RequestedFig}
\end{center}
\end{figure}

\newpage

\begin{figure}[!h]
\begin{center}
\includegraphics[scale = 0.5,trim =2cm  0cm 0cm 0cm]{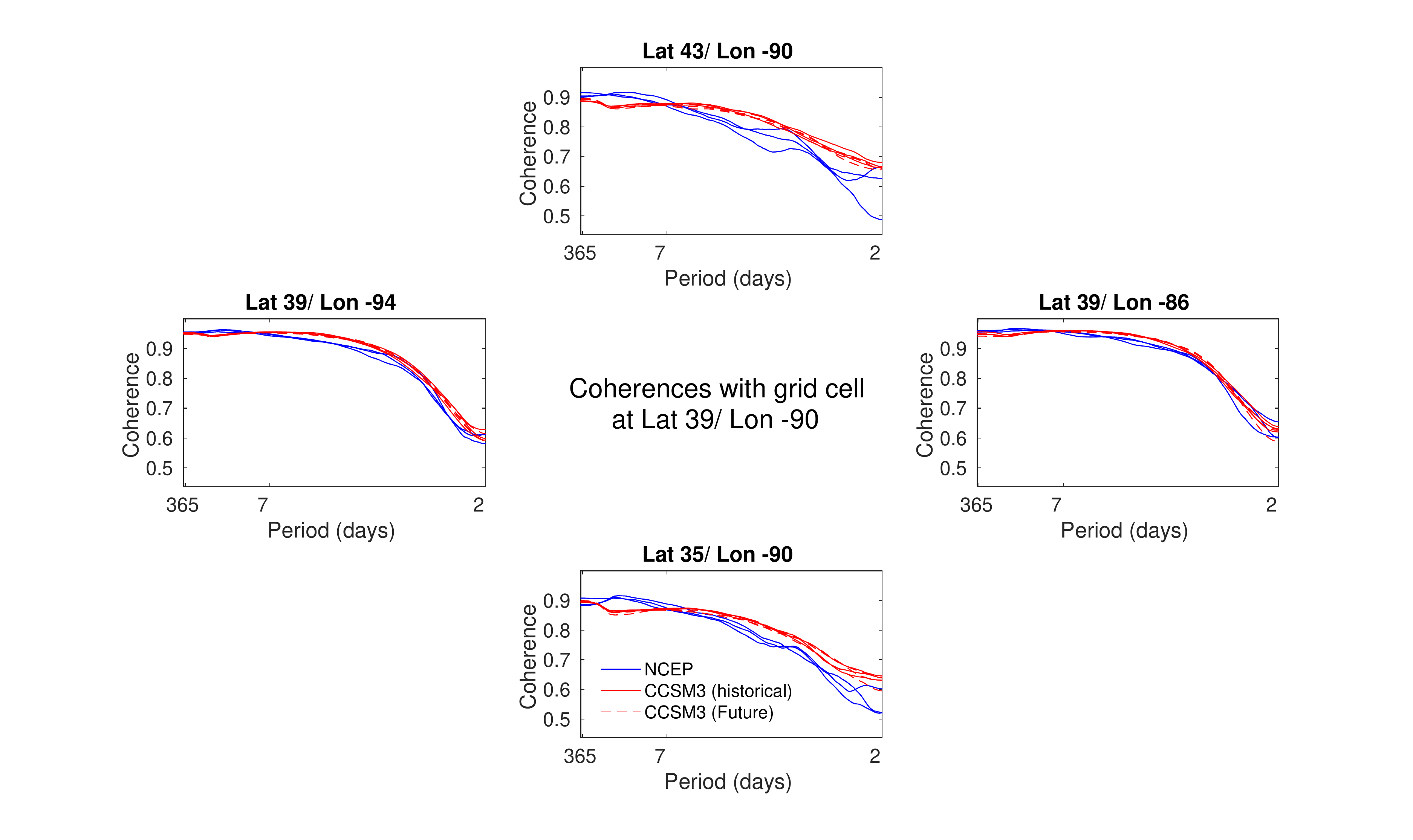}
\caption{Estimates of coherence spectra between temperatures in the Midwestern grid cell discussed in Section 5.1 and its four neighboring grid cells; the coordinates corresponding to each neighbor are indicated (and figures are positioned geographically). Estimates are shown for the ten-year blocks 1980-1989, 1990-1999, and 2000-2009 as well as the corresponding time 90 years in the future under the high scenario (i.e., 2070-2079, 2080-2089, and 2090-2099). Estimates in red correspond to temperatures from CCSM3 under historical forcing (solid) and in the high scenario future (dashed). Those in blue correspond to temperatures from NCEP CFSR and the simulation (the coherences do not change in the simulation). The time series were processed to remove means and (marginal) seasonal variability as described in the main text. Marginal spectra, co-, and quadrature spectra were estimated by averaging the raw estimates for each run (for the CCSM3 runs, where we have multiple runs) and then smoothing each component using a quadratic kernel; since we cannot average over multiple runs of reanalysis, the estimates for reanalysis were smoothed using a slightly wider bandwidth. While the coherence spectra are arguably similar in the East-West direction, CCSM3 has temperatures that are more coherent than the reanalysis's in the North-South direction. Coherences do not change much in the future GCM run, which suggests that our approach for the simulations may be adequate.}
\label{fig:cohPlotIL}
\end{center}
\end{figure}

\newpage

\begin{figure}[!h]
\begin{center}
\includegraphics[scale = 0.5,trim =2cm  0cm 0cm 0cm]{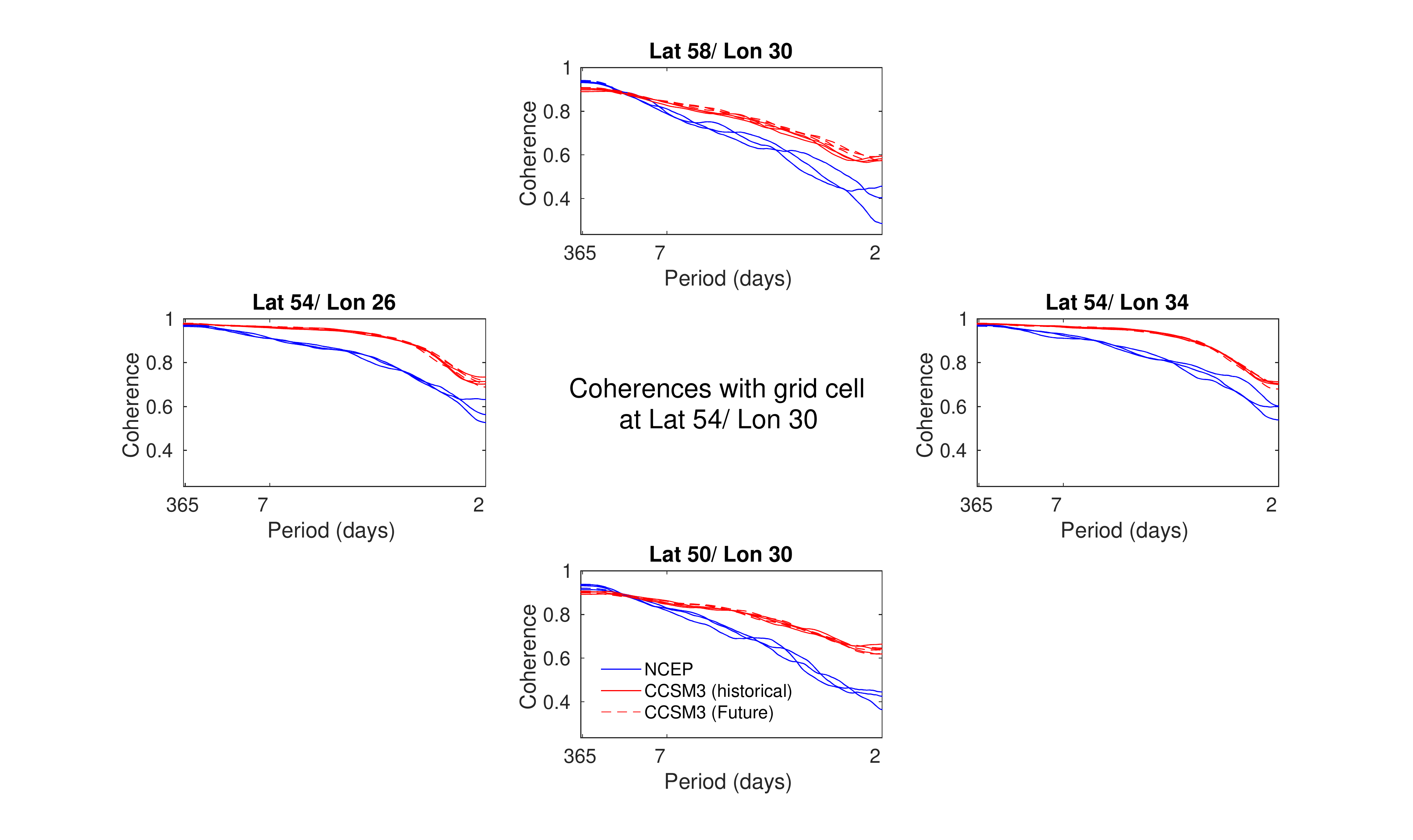}
\caption{The same as Figure~\ref{fig:cohPlotIL} but for a central grid cell in Northern Europe. Temperatures are more coherent in CCSM3 than in the reanalysis.}
\label{fig:cohPlotNE}
\end{center}
\end{figure}

\newpage

\begin{figure}[!h]
\begin{center}
\includegraphics[scale = 0.5,trim =2cm  0cm 0cm 0cm]{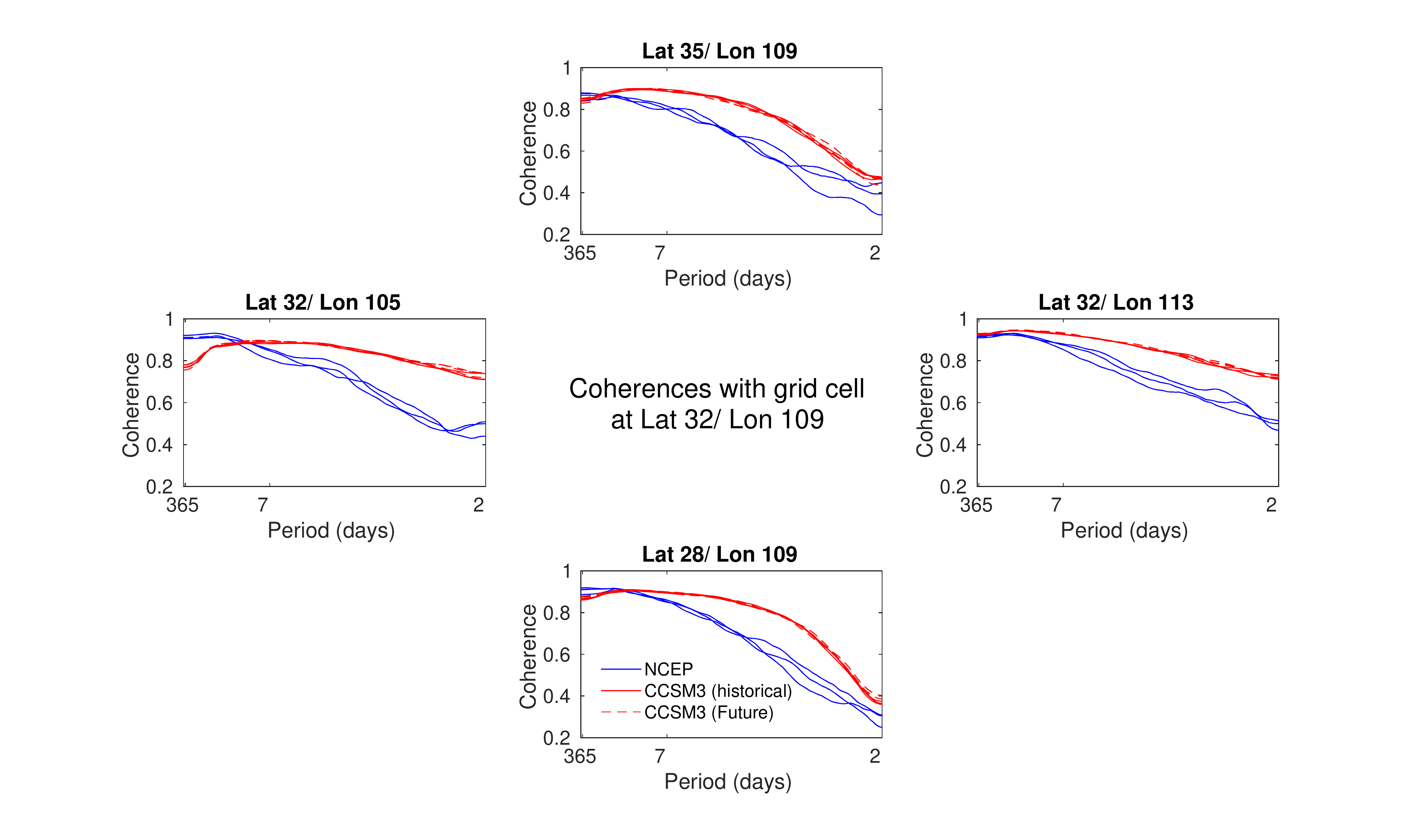}
\caption{The same as Figure~\ref{fig:cohPlotIL} but for a central grid cell in East Asia. Temperatures are more coherent in CCSM3 than in the reanalysis.}
\label{fig:cohPlotEA}
\end{center}
\end{figure}

\newpage

\begin{figure}[!h]
\begin{center}
\includegraphics[scale = 0.5,trim =2cm  0cm 0cm 0cm]{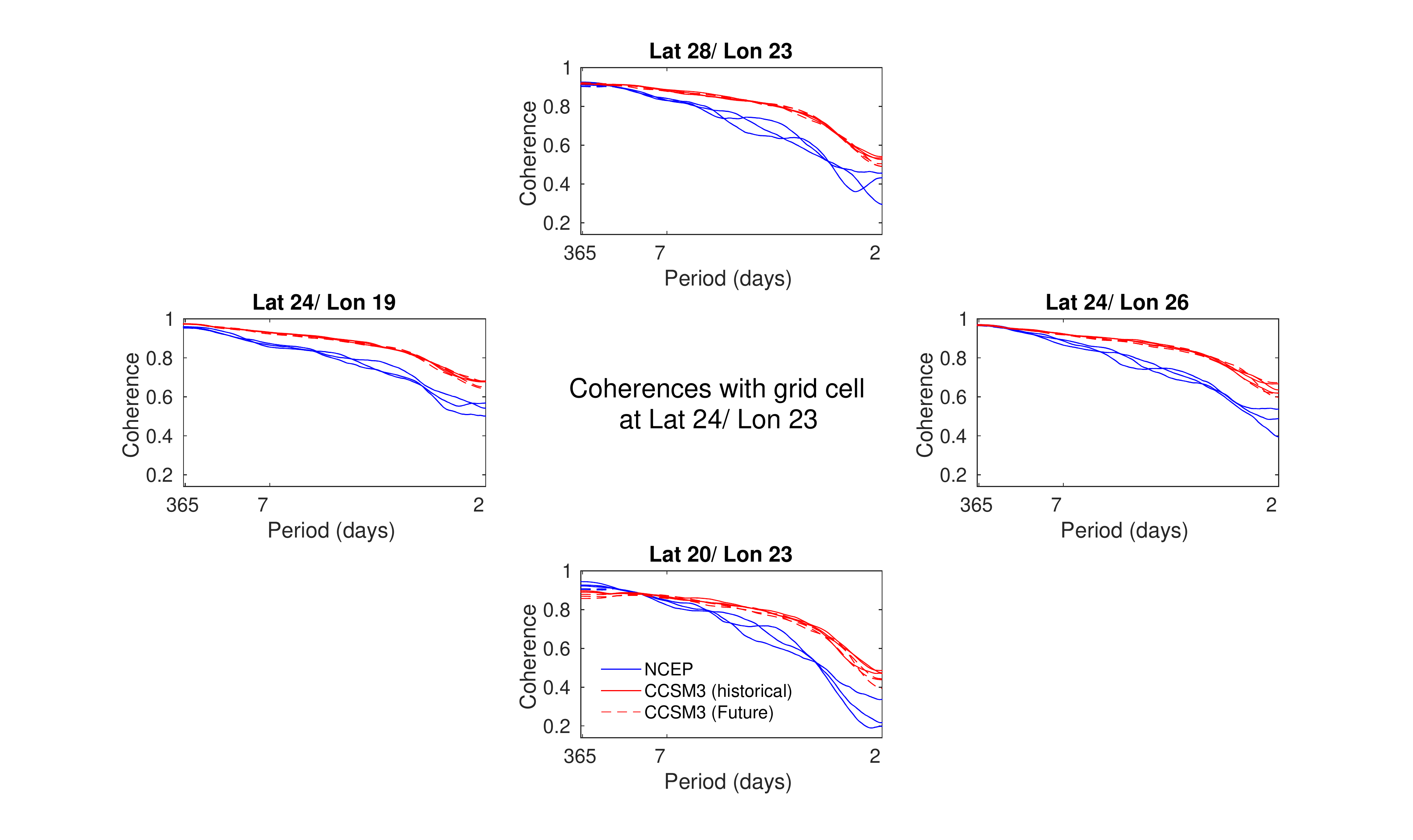}
\caption{The same as Figure~\ref{fig:cohPlotIL} but for a central grid cell in the Sahara. Here temperatures are more coherent in CCSM3 than in reanalysis at the higher frequencies.}
\label{fig:cohPlotS}
\end{center}
\end{figure}

\newpage

\begin{figure}[!h]
\begin{center}
\includegraphics[scale = 0.25,trim =0cm  0cm 0cm 0cm]{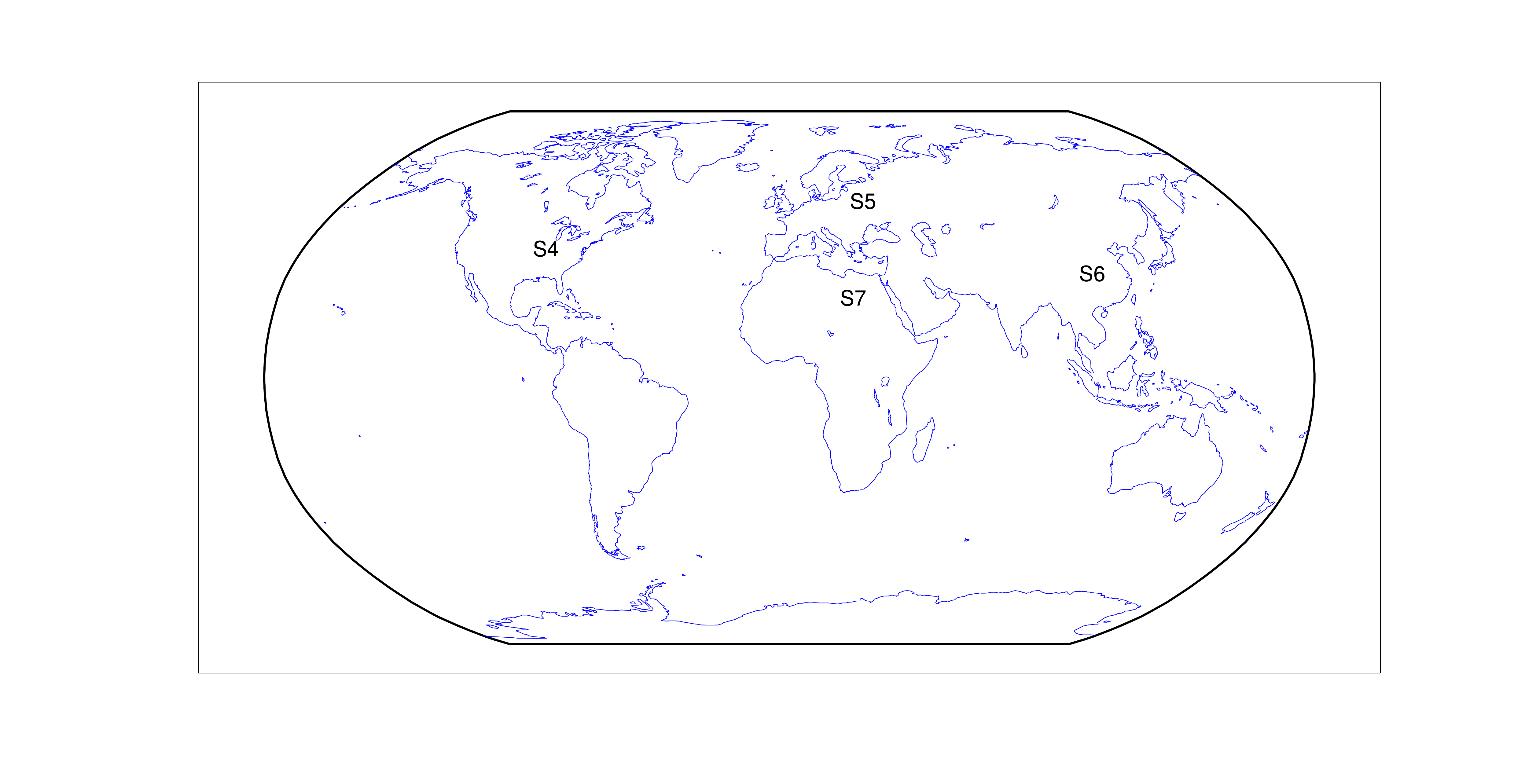}
\caption{The locations of the central grid cells corresponding to the coherency plots in Figures~\ref{fig:cohPlotIL} - \ref{fig:cohPlotS} (labeled by figure number). The grid cell for Figure~\ref{fig:cohPlotIL} corresponds to the one discussed in Section 5.1 of the main text.}
\label{fig:cohLocs}
\end{center}
\end{figure}

\end{document}